\documentclass{article}
\usepackage[T1]{fontenc}
\usepackage{lmodern}
\usepackage[ansinew]{inputenc}
\usepackage{graphicx}
\usepackage{epsfig}
\usepackage{bbm}

\begin{document}
%\draft

\title{The Quantum Hall Effects: Philosophical Approach }

\author{P. Lederer\\
Laboratoire de Physique des Solides, Universit\'e Paris-Sud et CNRS\\
Campus d'Orsay, 91405 Orsay Cedex, France\\
 pascal.lederer@u-psud.fr; 33 (0)6 62 98 40 51
 }
%\address{IPHST, Universit\'e Paris 1 }

\maketitle

\begin{abstract}
The Quantum Hall Effects offer a rich variety of theoretical and experimental advances. They provide interesting insights on such topics as gauge invariance, strong interactions in Condensed Matter physics, emergence of new paradigms. This paper focuses on some related philosophical questions. Various brands of positivism or agnosticism are confronted with the  physics of the Quantum Hall Effects.  Hacking's views on Scientific Realism, Chalmers' on Non Figurative Realism are discussed. It is argued that the difficulties with those versions of realism may be resolved within a  dialectical materialist approach. The latter is argued  to provide a rational approach to  the phenomena, theory and  ontology of the Quantum Hall Effects. 
\end{abstract}
\section{Introduction}
 Bachelard \cite{bachelard} stresses that  Philosophy must submit to the teachings of Science. As a physicist and a  philosopher of science, I am inspired by this point of view. In  the following, I am introducing a study of a relatively new field of physics, the Quantum Hall Effects, and trying to extract some relevant  philosophical view point from that study. 

The structure of the paper is as follows: the first parts of this paper (sections 2 to 5) are devoted to an  elementary introduction to this field \footnote{Some readers will find that this introductory part is not so elementary, as it requires some training in quantum mechanics... Is it conceivable to deal with the philosophy of physics nowadays  without a sufficient amount of knowledge of quantum mechanics?}.
 
Section \ref{progress} sets the historical stage which allowed for the appearance of the Quantum Hall Effects (hereafter QHE). Section \ref{revo} explains why the QHE qualify as a scientific revolution. Section \ref{history} specializes in the  history of these effects, beginning with the classical one, and introducing some simple  elements of theory for the motion of electrons in a magnetic field. Subsections \ref{classic}  and \ref{LL} deal, respectively, with the classical Hall effect, ancestor of the QHE, and the basic  quantum theory  of electron dynamics in a magnetic field. Subsection \ref{prelude} describes the ``normal science'' prediction, published before experiments were conducted. The astonishing experimental discovery of the Integer QHE \cite{IQHE} (hereafter IQHE), which seemed to refute qualitatively those  theoretical expectations, is described in subsection \ref{klitz}. Subsection \ref{storm}  introduces the second revolutionary finding: the Fractional QHE (hereafter FQHE) \cite{FQHE} and briefly introduces the  discoveries for which Laughlin \cite{laughlin1,laughlin2}  is responsible, such as fractional statistics and fractionally charged excitations; a novel theoretical entity, Composite Fermions, is mentioned. It was introduced by Jain  \cite{jain} as a development of Laughlin's theory, to account for some experimental results which the latter did not explain. Section \ref{complem} enters in more technical details, while remaining at a simple pedagogical level. Subsection \ref{integer} summarizes the main points of the theory for the IQHE based on Laughlin's work \cite{laughlin1}. Subsection \ref{gauge} discusses some aspects of gauge symmetry which are relevant in the FQHE theory. Subsections \ref{topological} and \ref{ferro} introduce two new theoretical  and experimental entities which are  by-products of the QHE: the topological insulator and the Quantum Hall ferromagnet. Section \ref{FQH} is devoted to the second Quantum Hall revolution: the experimental  discovery \cite{FQHE} and  the theory of the Fractional Quantum Hall Effects (hereafter FQHE)  also  by Laughlin \cite{laughlin2}; subsection \ref{CF} explains the main idea at the basis of  the Composite Fermion proposal.
%%%%%%%%%%%%%%%%%%%%%%%%%%%%%%%%%%%%%%%%%%%%%%%%%%%%%%%%%%%%%%%%%%%%%

 In the last parts, (section \ref{philo}) I will attempt  to draw some philosophical inferences from the material described in the previous sections.  In particular, I will discuss two versions of  "`Realism"'. First, I will  spend some time discussing "Scientific Realism", as developed by Ian Hacking (section \ref{sr}),  in his book "{\it{Representing and Intervening, Introductory topics in the philosophy of natural science"}} \cite{hacking}.  Hacking's views on the importance of practice in establishing truths about  the world will be stressed; this section contrasts Hacking's views with those of dialectic materialism. Section \ref{truth} discusses the possibility  of establishing truths about the world from Hacking's point of view, in relation with the QHE. This bears also on the   Non Figurative Realism picture  developed by Chalmers in his book on ``{\it{What is this Thing Called Science?}}''\cite{chalmers} which is mentionned in section \ref{NFR}. Scientific pluralism is  discussed in section \ref{pluralism}; section \ref{about} discusses the QHE from the point of view of various other  science philosophers. In section \ref{contradict},  I discuss some of the relations this study may have with the question of the unity and struggle of opposites in nature, namely one of the thesis of dialectic materialism in Nature. 

The conclusion lists the main results of this work (section \ref{conclu}).

The discoveries  of the IQHE in 1980 , and of the FQHE   in 1982, deal with an apparently restricted class of quantum phenomena: the behaviour of electronic systems in a two dimensional space under strong magnetic fields perpendicular to the two dimensional sample \cite{sarma}. What could we possibly learn about nature or about knowledge  which could be of any universal interest?

This paper aims at offering some answers to this question.  

\section{The result of theoretical and experimental progress}\label{progress}

A first  observation is that the developments which are the topic of this paper were made possible by progress in the physics of semi-conductors. The latter is an intimate mixture of theoretical and experimental progress, based in particular on the quantum mechanics of electrons in various pure and impure crystalline structures.

At the interface of two types of semi-conductors, experimentalists have been able to create electron populations which are confined, at low enough temperatures, to a thin spatial slice of the order of a nanometer. This is made possible by mastering  the theory and experiments on the  electronic band structure of the relevant semi-conductors, and of their interface: the energy for an electronic excitation to migrate to positions far from the interface can be made  a few orders of magnitude larger than  temperatures of order 100 K, while the energy for an electronic displacement in the interface is much smaller. Then, at low enough temperatures, electrons are restricted to a  two dimensional (2D) world.  This, in turn, was made possible by advances in the purity and regularity control of the crystalline arrays at the interface. The reader will notice how important is the notion of ``order of magnitude'': of energy compared with temperature, of distances compared to interatomic ones, etc. This notion was first highlighted in the philosophy of physics, I believe, by Bachelard \cite{lecourt}.
% As will be discussed in the last sections, this notion is essential in the debate of what we can call an absolute truth in physics. It will appear many times in this paper.  Another  catchword here is "mobility".

 Improving the mobility of electrons at the interface of specially selected semi-conducting materials was a precondition for the experimental and theoretical study of quantum particles in a two dimensional environment\footnote{At the time of this writing (april 2014), the electronic mobility at the interface of so-called GaAs/AlGaAs heterostructures is larger than $5. 10^7 cm^2/Vs$, which allows to observe details at a much finer scale than at the time of the original discovery of the IQHE}. This sets another example of the importance of the order of magnitude of the entity under study, in comparison with orders of magnitude of other relevant entities. The concept "`order of magnitude"' is a relational one.

%Another decisive progress in the story of Quantum Hall Effects was the advance in the technology of magnetic fields. Mankind is still unable to manufacture magnetic %field intensities larger than about $100$ Teslas. The typical electronic energies involved with the coupling to magnetic fields do not exceed 1 degree Kelvin per %Tesla, so that  temperatures significantly  lower than  $100$ degrees Kelvin  are usually required to study the effect  of the magnetic field on electrons as is %discussed in this paper.

Since the discovery of the QHE, two very different experimental systems have been found to exhibit QHE. An anisotropic crystal of weakly coupled  organic  conducting filaments  exhibits the IQHE \cite{HLM}; then the discovery of graphene \cite{novoselov} in 2004 has provided a genuinely two dimensional electronic system: a sheet of Carbon, the thickness of an atomic radius, can be sliced off a graphite crystal. Although I will not discuss those systems in this paper, it is worth mentioning that electrons in graphene have zero inertial mass, obey a relativistic Dirac equation, and move in a two-dimensional space where the ``velocity of light''  is the Fermi velocity, two to three orders of magnitude slower than the actual velocity of photons in the vacuum.

\section{A  revolution in theory} \label{revo}

The observation of Quantum Hall Effects, and particularly that of FQHE, participated in changing significantly the theoretical outlook on electronic liquids in condensed matter physics. Indeed the collective behaviour of electrons in simple metals\footnote{For example elements in the 3d and 4d transition series in the periodic table of elements.} has been described with considerable success by the Landau liquid theory \cite{landau}. Within that picture, interactions between electrons alter {\it{ adiabatically}} the ground state, as compared to that for which  interactions are neglected. In other words interactions in that picture are considered as perturbations which modify only quantitatively the parameters of the theory which has no interaction: no spontaneous symmetry breaking is usually expected. From a technical point of view, the main theoretical methods used in this context have been those of Feynman diagrams, and quantum field theory. 

Superconductivity has been explained, in part, with the Landau liquids as starting point \footnote{The term ``Fermi liquid'' is also used in this context}, and the mechanism for its breakdown -- together with a spontaneously broken gauge symmetry -- with the BCS theory: at low temperatures, the Fermi sea becomes unstable to the formation of electron pairs, due to an effective attractive interaction mediated by the vibration quanta of the crystalline network \cite{BCS}; those pairs are bosons which (to be simple)  form a Bose-Einstein condensate. This theoretical framework, although responsible for some irreversible advances in the understanding of a large class of insulators, semi-conductors and conductors\footnote{So called large band electronic  systems.}, has met its limits with the discovery of the QHE, and of a number of other phenomena, such as Mott insulators, or  the high temperature superconductivity in copper oxydes in 1986 \cite{bednorz}.

New theoretical methods have been necessary to account for such discoveries, and in particular that of the QHE. The path followed by Laughlin \cite{laughlin1,laughlin2} led him to the Nobel prize award in 1993. What is revolutionary in the QHE  is : a) the original Laughlin's method  to find  the almost exact ground state  wave function for strongly interacting fermions (electrons) in two space dimensions, on the basis of symmetry considerations, within a particular gauge choice; b) this opened -- within the vast field of quantum mechanics --  a new field of physics, with new methods, new theoretical entities, such as incompressible quantum fluid, Composite Fermions, or topological insulators. The method was  a radical departure from  methods that had been followed until then.

In perturbative approaches, the theorist starts from a known solution for the non interacting problem, which is related to  the single particle problem.  Although this approach may seem to justify a reductionist point of view (i.e. the properties of electron liquids would be the sum of properties of single electrons), the statistical  properties of electrons, due to the antisymmetry of their wave function, introduce from the start a major correction (the exclusion principle) to a naive reductionist viewpoint.

A useful way to think about many aspects of physical systems is that they embody conflicting energies: in the case of electronic systems, the conflict between kinetic energy and interaction energy is often essential. The reason why they are in conflict is that the equilibrium properties, which govern  a large number of physical properties, depend on reaching an energy minimum\footnote{In thermodynamics, for a system in contact with a thermostat, the two energies in conflict are the internal energy, on one hand, the entropy term, on the other hand. The conflict is then between order and disorder, which interpenetrate each other, when symmetry breaking is involved.}. The configuration of particles, or the density, which minimizes, at equilibrium,  the kinetic energy is in general qualitatively different from that which minimizes their interaction energy.  When the kinetic energy dominates, the wave functions spread in space as much as is possible; repulsive  interactions  in this case are a weak  perturbation of the system with zero interaction: on the theory side,   a zero order solution is easily constructed from the knowledge of the single particle behaviour.

 When interactions dominate, each particle becomes an obstacle to the motion of all others; the hopefully simple limit of infinite interactions is in general one with a ground state which  has a formidable degeneracy, and there is no intuitive way to guess how this degeneracy may be lifted when interactions become finite. A whole new world of symmetry breaking possibilities opens up, and the reductionist point of view is of little help: the properties of the whole may differ qualitatively from that of its parts. This is at the basis of so-called  emergent properties. 

%%%%%%%%%%%%%%%%%%%%%%%%%%%%%%%%%%%%%%%%%%%%%%%%%%%%%%%%%%%%%%%%%%%%%%%%%%%%%%%%%%%%%%%%%%%%%%%%%%%%%%%%%%
What Laughlin showed is that, in the case of the Quantum Hall Effect,  physical intuition,  a careful analysis of the symmetries of the problem, and clever choice of theoretical language, could provide for an answer, at little or no computational cost\footnote{P. W. Anderson \cite{PWA} had, some years before the nineteen eighties, devised a theoretical proposal for a different electronic  system: the Resonating Valence Bond paradigm, which is outside the scope of this paper.}. In most problems so far dealing with many body physics, after the nineteen fifties, quantum field theory had been considered a necessary tool.  Laughlin broke away from this paradigm and provided an answer written in terms of bona fide wave functions, written in first quantization language\footnote{In time, his theory was cast  in the quantum field theory language.}. The choice of theoretical language, namely quantum field theory or first quantization, seems, at first sight, to be  a matter of convenience, at least in non relativistic physics, which is in general the case of condensed matter physics. Quantum field theory is a technical progress in  handling many-particle states, because of their symmetry (bosons) or antisymmetry (fermions) under permutation of particles.  Contrary to some philosophy papers, quantum mechanics of many particle systems do not alter the fundamental duality/conflict of waves and particles, which is at the basis of all quantum formulations, either in terms of quantum fields or in terms of  first quantized  wave functions: the Schr$\ddot{o}$dinger equation is the fundamental starting point.

\section{History of the Quantum Hall Effects}\label{history}

\subsection{Prehistory: the classical Hall effect}\label{classic}
\begin{figure}
\centering
\includegraphics[width=10.5cm,angle=0]{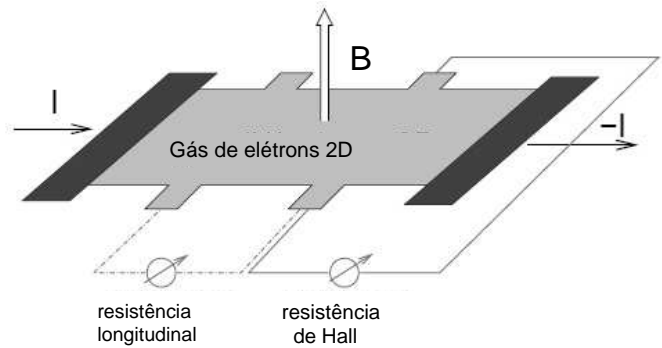}
\caption{\footnotesize{ two dimensional electron system in a perpendicular magnetic field B. A current $I$ is injected through the black contacts. A longitudinal resistance is measured between two contacts on the same edge of the metallic bar; the Hall resistance $R_H$ is measured between contacts on opposite edges.}}
\label{fig01}
\end{figure}

The classical Hall effect was described by Hall in 1879 : if an electrical current is driven through a flat metallic bar (see figure (\ref{fig01})) in the presence of a static magnetic field $B$ the direction of which is orthogonal to the bar surface, a gradient of electronic density is created in the bar in a direction orthogonal to that of the current. This gradient leads to a potential difference $V$ between the edges of the bar. This defines a "Hall resistance" $R_H$ which is proportional to the field and inversely proportional to the electronic density, $R_H= -\frac{B}{en_{el}}$. $n_{el}$ is a surface density if the current is stricly two-dimensional and $-e$ is the electron charge. The classical theoretical graph  for $R_H$ is thus a straight line  as a function of $B$, with a slope  inversely proportional to the electron density.

The experimental set-up sketched in figure \ref{fig01} is a perfect example of what Bachelard \cite{bachelard} called a {\it{phenomenon operator}}  i.e. an artifact constructed in the laboratory, on the basis of theoretical views, to investigate the response of nature to specific stimulations. Hacking \cite{hacking} developed similar views on experimental set-ups.

In a perpendicular magnetic field, the classical electron motion in free 2D  space is circular, with frequency $\omega_C =eB/m$.

 In the presence of an electric field $\vec{E}$ and  a magnetic field $\vec{B}$, the straightforward analysis of the resistivity, when interactions can be neglected, relies on the (Newton) equation of motion of a single classical charge\footnote{In this context this is called the Lorentz force equation.} which describes the time evolution of the particle  momentum $\vec{p}$: 
\begin{equation}
\frac{d\vec{p}}{dt} = -e\left( \vec{E} + \frac{\vec{p}}{m}\wedge \vec{B}\right)
\end{equation}
This equation is valid when  dissipation processes are neglected. 
The current density is defined as $\vec{j}=-n_{el}e\vec{p}/m$. The resistivity $\rho$ is defined by the (matrix) equation
\begin{equation}
\vec{E} =\rho \vec{j}
\end{equation} 
This defines longitudinal resistivity by, for example, $E_x=\rho_{xx}j_x$, and transverse resistivity by $E_x= \rho_{xy}j_y$ .

The result is:
\begin{equation}\label{ro}
\rho=\left(\begin{array}{cc} 0 & \displaystyle\frac{B}{en_{el}}\\
- \displaystyle\frac{B}{en_{el}}& 0 \end{array}\right) \hspace{2cm}
\sigma = \left(\begin{array}{cc} 0 & -\displaystyle\frac{en_{el}}{B}\\
\displaystyle\frac{en_{el}}{B} & 0 \end{array}\right).
\end{equation}

Thus $\rho_{xx}= \rho_{yy}= \sigma_{xx} = \sigma_{yy} =0$ and the Hall resistivity $\rho_{xy}= -\rho_{yx}$.

The conductivity matrix $\sigma$ is the inverse of the resistivity matrix: $$\sigma = \rho^{-1} = \left(\begin{array}{cc} \sigma_L & -\displaystyle\sigma_H\\
\displaystyle\sigma_H & \sigma_L \end{array}\right).$$
Notice that the result (\ref{ro}) is counter-intuitive: the  longitudinal conductivity vanishes together with the longitudinal resistivity. This is due do the transverse character of transport in a magnetic field, which is also the reason for the matrix formulation.

\subsection{Quantum mechanics -- Landau quantization}\label{LL}

In the nineteen thirties, Landau studied the quantum mechanical motion of an electron in a magnetic field in two dimensions. He found that the equations of motion are those  of a quantum harmonic oscillator, so that, as for the quantum oscillator, the energy levels (which in this problem are called "Landau Levels") of the electron are equidistants, with a spacing $$\hbar \omega_C=\hbar\frac{eB}{m}.$$  The energy levels $\epsilon_n$ are labeled by positive integers $n$: $$ \epsilon_n = \hbar \omega_C (n+1/2). $$ Each level has a large field dependent degeneracy: the number of states per Landau Level (hereafter noted LL) is $N_B=n_B A$, where $A$ is the sample surface, and $n_B= B/\phi_0$ is the density of magnetic flux through the sample surface, measured in units of flux quantum $\phi_0 =h/e$. Typically, for a magnetic field of order a few Tesla, $n_B \approx 10^{11}$ par cm$^2$, a large degeneracy indeed.

Since electrons are fermions, they gradually fill the LL electronic states, as their number increases. It is natural to define a LL filling factor $\nu$:
\begin{equation}
\nu = \frac{n_{el}}{n_B}
\end{equation}
\begin{figure}
\centering
\includegraphics[width=6.5cm,angle=0]{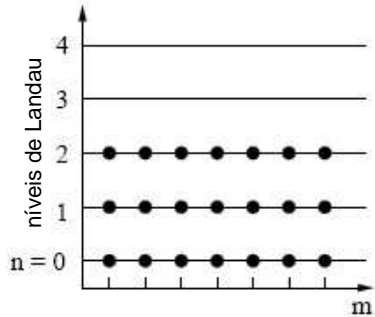}
\caption{\footnotesize{ Landau Levels. The quantum number (integer) $n$ labels the levels, and $m$ is associated to the center of motion which governs the degeneracy  of each level.}}
\label{fig1:02}
\end{figure}
 Figure (\ref{fig1:02}) is a schematic diagram of the Landau Levels and their occupation by single electron states.

Interesting  magnetic effects -- well known in 1980 -- occur at low magnetic fields when the number of occupied Landau Levels is large; I do not discuss them in this paper. The Quantum Hall Effects arise at large  fields, such that the LL degeneracy is of the order of the number of electrons in the sample, i.e. such that only a few Landau Levels are occupied.

\subsection{The ``normal science'' prediction: the Wigner Crystal.}\label{prelude}

As mentioned above, theoretical and technological advances in the nineteen seventies  allowed to foresee the experimental  realization of  two-dimensional samples with good control of the electron density. The experimental realization of a situation where only one Landau Level (the lowest energy one) would be partially or completely occupied became a practical possibility. It was natural to explore the theoretical prediction for the behaviour of such electron systems under   perpendicular magnetic field. The Coulomb interaction energy has to be much larger than the kinetic energy, which practically vanishes from the picture when the lowest LL only is completely or partially occupied. The ``normal science'' prediction (to use  Kuhn's language) was to predict the occurrence, when only the lowest Landau Level is occupied,  at low temperature of a ``Wigner crystal'' or, equivalently, a ``Charge Density Wave'' \cite{FPA}. The former  is a compressible crystalline array of electrons stabilized by the minimization of the interaction energy: instead of organizing in a Fermi sea of itinerant electrons with overlapping wave functions which spread over the whole sample volume, the electrons stay localized  around lattice points;   staying as far apart from one another as possible minimizes their interaction energy. Such objects are known to exist in other areas of Condensed Matter physics. A straightforward prediction, associated   to this idea, was that this would result in an insulating behaviour below the temperature for crystallization of the electrons. This again was a well known logical consequence of the inevitable sample defects (impurities, dislocations, etc.) which are known to pin the electron crystal to the sample atomic lattice, thus suppressing the electronic conductivity. Another straightforward prediction was that this Wigner Crystal, which breaks the continuous translation symmetry of a liquid, would possess phonon modes, i.e. quantized vibrations of the electron lattice, with an energy mode going continuously to zero with their inverse wavelength.

In other words, the prediction was that below the Wigner crystal crystallization temperature, the  longitudinal and  Hall resistance   would diverge exponentially with inverse temperature, and  low energy phonon modes would exist. 

\subsection{The Integer Quantum Hall Effects} \label{klitz}
%%%%%%%%%%%%%%%%%%%%%%%%%%%%%%%%%%%%%%%%%%%%%%%%%%%%%%%%%%%%%%%%%%%%%%%%%%%%%%%%%%%%%%%%%%%%%%%%%%%%%%%%%
 
\begin{figure}
\centering
\includegraphics[width=13.5cm,angle=0]{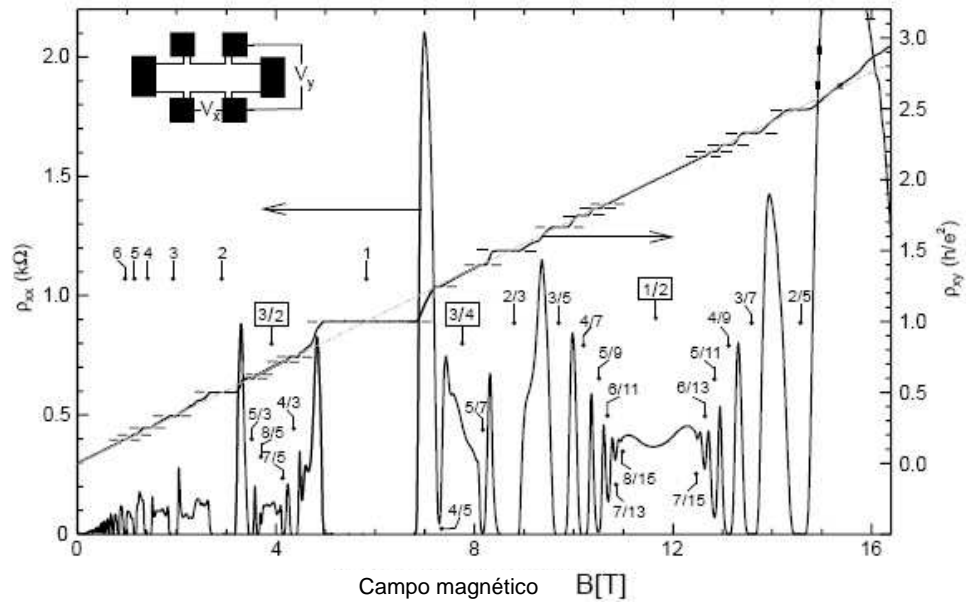}
\caption{\footnotesize{ Experimental signature of the quantum Hall effect. Each plateau in the Hall resistivity $\rho_{xx}$ coincides with a zero longitudinal resistance. The classical curve expected for the Hall resistance is the dotted line. Numbers indicate the filling factor $\nu=n$ for the IQHE, and $\nu= p/q$ ($p$ e $q$ integers) for the FQHE. The integer QHE is observed for $B$ smaller than about 8 Tesla, the fractional QHE at larger B values. The inset in the upper corner left is  a schematic view of the Hall effect experiment.}}
\label{fig02}
\end{figure}

The classical Hall resistance is represented on figure (\ref{fig02}) by a dotted line.

Von Klitzing  undertook the task to check the Wigner Crystal prediction with the available samples and magnetic fields.
 His experimental project was, as usual, heavily intertwined with the theoretical notions described above. 

His findings were radically different from the predicted effect.

 Not only did the   Hall (transverse) resistance develop plateaux around  integer values of the filling factor of electronic LL, but the longitudinal resistivity was found to vanish exponentially with temperature around the same values! There was no sign of any tendency towards a compressible insulating Wigner Crystal. And there is no sign of low energy phonon modes. On the contrary, a gap is observed to  govern the low temperature properties.
 
Furthermore, the conductivity for each plateau appeared to be a constant (identical for all plateaux) times an integer, the number of totally filled Landau Levels, which characterizes each plateau.

The ``normal science'' prediction was a total failure.
% If one believes Popper or  Kuhn \cite{kuhn,popper}, this might have led to a complete change of paradigms.
 Should physicists have given up Fermi statistics, or the quantum description of electronic motion in the presence of magnetic  or electric fields, or perhaps quantum mechanics? Should they have  regarded all or part of those theories as obsolete beliefs? None of this happened.
% confirming Lakatos' views on Popper's falsification theory.
 The Wigner Crystal prediction was (temporarily) abandoned, and the experimental result acquired the status of riddle.

Just as the failed Wigner Crystal prediction, the IQHE discovered by v. Klitzing \cite{IQHE} in 1980 (Nobel prize in physics in 1985), is in fact a  consequence of Landau Level quantization, together with the Pauli principle obeyed by fermions. In contrast with the failed theoretical predictions, Laughlin \cite{laughlin1}  concentrated first on electron transfer from one edge of the Hall bar  to the other.
%, which he showed were smaller than the interlevel energy difference $\hbar\omega_C$
 %One important feature which allowed to account for the IQHE  within the previously known laws of physics is the unsuspected role of sample boundaries.

The IQHE is due to an incompressible electronic liquid state. Incompressibility here means there is a gap between the ground state and the first excited energy level in the structure of energy levels of the system. At low temperatures, when $k_BT<< \hbar \omega_C$,  constant Hall resistance plateaux appear; they are quantized as $$R_H=\frac{h/e^2}{n}$$ where $n$ is an integer,  the integer part of the LL filling factor: $n=\left[\nu\right]$. The proportionality constant is a universal constant $h/e^2$! Each plateau in the Hall resistance coincides with zero longitudinal resistance, or, more precisely, exponentially decreasing $\rho_{xx}$ with temperature ( low field part of figure (\ref{fig02}).

In fact, the Wigner Crystal hypothesis  was subsequently proved correct, but  in a different range of parameters, and found to  describe   the physics of the 2D electron system under magnetic field at lower filling fraction\footnote{Initially, Laughlin had incorrectly estimated the filling  fraction below which the Wigner Crystal would become the stable ground state; he found the boundary at $\nu \approx 1/70$.} $\nu \leq 1/6.5$, and in certain cases for higher  LL quantum numbers than those for which various QHE states are stabilized. In particular, new details of the Hall resistance experiments have been explained by a competition of QH states with a novel version of the  Wigner Crystal \cite{goerbig}:  crystalline arrays of   electron droplets, each droplet containing an integer number (larger than one)  of electrons. 
%This development is yet another argument against Popper's falsification theory, in agreement with  criticisms  by Feyerabend and Lakatos (\cite{feyer,lakatos}): indeed 

This last paragraph may sound confusing. How can the Wigner Crystal prediction of reference \cite{FPA} be wrong in a given range of LL filling fraction, and right in another? The explanation is that the free energies for the Wigner Crystal and that of the QH state compete with one another, and the competition depends on the LL filling fraction $\nu$. Whichever has the lowest free energy  in a given range of $\nu$ values will be stabilized at the expense of the other. The authors of the Wigner Crystal prediction \cite{FPA},  wrote their paper  before the experiment was done. It  revealed instead a new, unknown state of matter -- the QH state -- , because of its lower energy than that of the Wigner Crystal.  
%The Wigner Crystal proposal obeys all the known  laws of physics, and was in fault only because a lower energy solution exists, as Laughlin showed, in certain ranges of LL filling fraction. 

The Wigner Crystal  disagreement with experimental prediction in a certain range of $\nu$ values did not prove it wrong  as a possible physical state of the 2D electron system. As Laughlin proved  the QH states  had lower energy in the experimental range of parameters which was studied at first. In other ranges of LL filling fraction, the Wigner Crystal may be the winning physical state. 
%%%%%%%%%%%%%%%%%%%%%%%%%%%%%%%%%%%%%%%%%%%%%%%%%%%%%%%%%%%%%%%%%%%%%%%%%%%%%%%%%%%%%%%%%%%%%%%

The  resistance quantization  is independent of the sample geometry, which is a remarkable property in two spacedimensions: the Hall resistance $R_H$ at a plateau is given in terms of fundamental constants $e$ and $h$. The resistance value accuracy at $n=1$ is such ($\approx 10^{-9}$) that this plateau now serves to define the resistance metrology standard, the v. Klitzing constant, $R_K= 25812,807 $ ohms. In fact,  as appears from notes which von Klitzing scribbled on his experimental notebook, he immediately guessed that the value of the resistivity plateau he observed was the resistance quantum $h/e^2$.

\subsection{The FQHE} \label{storm}

The discovery of the FQHE \cite{FQHE}, which was another huge surprise,  followed in 1983. It occurs at "magical'' LL filling fraction\footnote{The term ``magical'' originates from the bewilderment of physicists in front of the rich plateau structure found experimentally, which baffled at first all expectations of what might happen at LL filling fraction smaller than $1$.} smaller than $1$, in particular that of the lowest LL. The first observed fractional plateau (i.e.  at fractional LL filling fraction) was at $\nu = 1/3$. Since then, with better samples, more accurate measurements, etc., a wealth of other plateaux have been observed. 

 Our basic understanding to-day is based on the FQHE theory  by Laughlin \cite{laughlin2} in 1983: he proposed  an almost exact  many particles  trial wave function  which describes an incompressible electronic liquid. In the FQHE as well, an incompressible quantum liquid is one  for which the excitation spectrum  is separated from the ground state by an energy gap.

 In this picture, Laughlin \cite{laughlin2} made the revolutionary  proposal that, associated with the fractional occupation of the LL, were various fractionally charged excitations with respectively $1/3, 1/5, 1/7...$ of an electronic charge.  This prediction has been later on proved experimentally correct, at least inasmuch as the existence of the $1/3$ fractionally charged excitation has been experimentally verified. Furthermore, such excitations obey ``fractional statistics''. The latter is a generalization of the statistics which govern quantum particles in three space dimensions; in the latter case, particles  obey either Fermi statistics (electrons, neutrons, protons) or Bose statistics ($^4$He, photons, $\alpha$ particles, etc.). In two dimensions, the exchange of two identical particles may multiply the wave function by a factor $\exp{i\theta}$ where $\theta$ can be {\it{any}} angle, whence the name {\it{anyon}} coined for such particles, the existence of which have overcome a long standing belief that particles have only a binary choice of statistics. In three dimensions, the angle $\theta$  for exchange of two particles is either $0$ (bosons) or $\pi$ (fermions).

Various generalizations followed. In particular the FQHE may well be the signature of the condensation --or emergence-- of a new type of particle, known as Composite Fermion \cite{jain} (hereafter CF). The latter are  bound states of an electron (a fermion) and an even number of flux quanta, a quite novel idea and a quite novel particle.  In fact Laughlin's wave function may be recast in the CF language. In this picture, the $1/3$  FQH plateau is a completely filled LL of CF. The surface per CF state is $(2s+1)$ times the surface of an electron state $2\pi l^2_B$.

   CF represent at least a useful theoretical tool to understand many new features of the Quantum Hall Effects. They have allowed theoretical predictions which have been successfully tested. 
	
	The history of physics has taught us that many new theoretical concepts which were thought at first to be mere mathematical constructs to account for phenomena ({\it{sauver les apparences}}, following Duhem \cite{duhem}), eventually were shown to reflect in a more or less detailed way objective  material features of the world. The heliocentric model of Copernicus, the atoms, the molecules, the electrons, the photons,   the molecular carriers of genetic information, etc., have transited in the course of history, from the status of useful hypothesis to that of more or less exact reflection of  real material entities. There are many theoretical and experimental facts which  suggest that the CF hypothesis may reflect our present knowledge of  such  real objects.
	
%%%%%%%%%%%%%%%%%%%%%%%%%%%%%%%%%%%%%%%%%%%%%%%%%%%%%%%%%%%%%%%%%%%%%%%%%%%%%%%%%%%%%%%%%%%%%%%%%%%%%%%%%%%%%%%%%%%%%%%%%%%%%%%%%%%%
Most advances in many-body physics,  until the discovery of the FQHE, had used perturbation methods, apart from Gell-Mann's eightfold way,  Mott's theory of a class of magnetic insulators, or Anderson's theory of the Resonating Valence Bond \cite{PWA}. What became clear with the FQHE was the unsuspected richness of strongly interacting systems, be they electronic or nuclear systems. The Yang-Mills gauge theories have developed gauge field theories  to study strong interactions, and spontaneous symmetry breaking, etc.. With his work on the FQHE, Laughlin chose a  different theoretical path, and showed that strong electronic interactions had in store many surprises, both theoretical and experimental. New concepts have emerged as the consequences of novel experimental phenomena, reflecting the existence of unforeseen particles, unforeseen realities. 

This is explained in more technical details below.

\section{The Integer Quantum Hall Effect}.\label{complem}

\subsection{Preliminary remarks; a different outlook on quantum complementarity: a by-product of the theory}
In the classical treatment of two dimensional electron motion in the presence of a (perpendicular)  magnetic field, it is found that electrons move in circular orbits characterized by a radius which depend on their kinetic energy. This is because the Lorentz force $e\vec{v}\wedge \vec{B}$ is orthogonal to the electron velocity $\vec{v}$, and does no work: it merely deflects the electron from a rectilinear trajectory in zero magnetic field. The coordinates $X$, $Y$  of the circle center are constants of motion, which may be located at any point on the sample surface.

The quantum treatment cannot start from the Lorentz force equation; it has to start with the Lagrangian or the Hamiltonian, which are formulated in terms of the potentials, in particular in terms of the vector potential $\vec{A}$, the curl of which yields the magnetic field. $\vec{A}$ lies in the plane perpendicular to $\vec{B}$, i. e. the sample plane. The details will not be given here. However an important remark is that gauge invariance enters the problem while it is absent in the classical Lorentz force equation. Gauge invariance proves an important tool \cite{laughlin1} for the flexibility of theoretical approaches of the quantum problem\footnote{See reference \cite{lederer} for a related discussion.}. 

The quantum treatment of the 2D electron system in perpendicular magnetic field also has  constants of motion $X$, $Y$ , i.e. operators which commute with the Hamiltonian $H$:
\begin{equation}
[X, H] =0 \hspace{2cm} [Y, H] =0.
\end{equation}
However a new feature appears: the well known non commutativity of the momentum operators $p_x, p_y$ with the position operators $x,y$ leads, for 2D electrons in the presence  of perpendicular magnetic field $B$, to  the following commutator of $X$ and $Y$:
\begin{equation}\label{XY}
\left[ X, Y \right] = il^2_B \hspace{2cm} \left[ \eta_x,\eta_y \right] = -il^2_B
\end{equation}
In the second  equation above, $\eta_x, \eta_y$ are the coordinates of the electron relative to $X,Y$. The length $l_B$ which appears is the so-called ``magnetic length'' defined by $l_B= \sqrt{\hbar/eB}$. The surface $ 2\pi l^2_B = \frac{h/e}{B}$ is that through which the flux of the field $B$ is one flux quantum $\phi_0=h/e$.

The commutator in equation (\ref{XY}) is, of course, gauge invariant.
% It shows that the quantum theory of the  2D single particle in a field is probably the simplest example of a non commutative geometry.
 In the zero field single particle quantum mechanics, the commutator of $p_x$ and $x$  defines a surface $2\pi \hbar = h$ in phase space, as found by Planck in his study of the black body radiation. Equation (\ref{XY}) defines a surface $2\pi l_B^2$ in the \underline{real  sample surface}. {\it{Real space in this problem plays the role of phase space in the zero field problem}}.  In other words each electronic state occupies in the sample a surface through which a flux quantum $h/e$ is threaded by the magnetic field.  The Heisenberg relations associated to the commutator in equation (\ref{XY}) are:
\begin{equation}\label{delta}
\Delta X. \Delta Y \geq 2\pi l^2_B.
\end{equation}
The surface $2\pi l_B^2$ plays the role that $h$ plays in the zero field quantum mechanics!  It is proportional to $1/B$. Typically for a field of a few Tesla, the length $l_B$  is of order of $10$ to $30$ nm.

Equation (\ref{delta}) expresses the fact that the  spatial extent of an electron state has   center of motion coordinates which  are spread over a surface $2\pi l_B^2$. In terms of a measurement language, it means that $X$ and $Y$ cannot be simultaneously determined with a better accuracy than $\Delta X, \Delta Y$ together with equation (\ref{delta}). 
%Depending of the gauge chosen to describe the sample, the accuracy on one coordinate may become very good, or very large, and is always related to the accuracy of the other coordinate %through (\ref{delta}).

Both uncertainty relations, i.e. equation (\ref{delta}), and the standard Heisenberg relations in the absence of field, are a direct consequence of the commutator  of $x$  and $d/dx$. However,  (\ref{delta}) deals with a surface in real space, in contrast to a surface in phase space. While the latter is an  abstract concept, the former is straightforward and concrete. This should allow the Heisenberg uncertainty principle to appear less mysterious to whoever wants to understand the meaning of quantum mechanics. Quantum mechanics in zero field sets a limit to the classical definition of a point in phase space, because the classical notion of trajectory -- given by the simultaneous determination of $p$ and $q$ -- becomes irrelevant at a microscopic level\footnote{A particle does not move along a single trajectory, but along a superposition of different trajectories.}. Similarly, equation (\ref{delta}) sets a limit to the classical  notion of center of motion  in a magnetic field: it is spread over the state  surface $2\pi l^2_B$. Infinite accuracy would  require an infinite field, for which $l^2_B$ would vanish,  which would  also mean infinite energy for the lowest LL. The complementarity of $X$ and $Y$ is straightforward, since their uncertainty serves to determine  a surface in real space: the surface occupied by a quantum state. Thus, the simplest quantum treatment of 2D electrons in a magnetic field, besides being a first step for the QHE theory, offers an interesting view point on complementarity.

The Landau Level degeneracy can be directly  found from the  surface associated to a flux quantum: the number of states per level and per unit sample  surface is obtained by dividing the latter by the flux quantum surface, i.e. $2\pi l^2_B$; or, equivalently, by dividing  the total flux per  surface unit  by the flux quantum: $n_B= B/\phi_0$. This result was mentioned above: it can be derived with no knowledge of the mathematical solution of the Schr\"{o}dinger equation, merely by using the commutation relations (\ref{XY}). 

A consequence of (\ref{XY}) is that a variation of $B$ induces a variation of the number of states available for electrons per Landau Level in a given sample, since the degeneracy of the latter is $Be/h$ per  surface unit. Since the electron density is in general fixed in a given sample, increasing the field intensity $B$ amounts, at temperature low compared to the inter-level energy distance\footnote{So that thermal excitations to higher energy levels are negligible.} $\hbar \omega_c$, to gradually emptying high energy levels of their electrons which are transferred to lower energy ones. What is happening when, at large enough $B$, the lowest Landau Level is only partially filled? The answer to this question was awarded the Nobel prize, and is described in section (\ref{FQH}).

\subsection{Basic ingredients of the IQHE}\label{integer}

The basic ingredients to understand the IQHE are:
\begin{itemize}
\item Each completely filled LL contributes a  conductance quantum $e^2/h$ to the sample Hall conductivity. {\it{The latter is thus defined in terms of universal constants $e$ and $h$!}}
\item Electrons added to, or subtracted from   the filled LL  are localized by the sample impurities, and do not contribute to electric transport, so that the conductivity remains pinned at its value for the filled LL, until electrons become delocalized. As mentioned above, increasing  the magnetic field intensity amounts to creating more states in each LL for a fixed number of electrons in the sample. Thus, starting from a filled LL situation, increasing the field intensity amounts to gradually emptying the highest occupied LL of its electron population. Conversely, starting from  a filled LL situation, and decreasing the field intensity, decreases the number of states per LL, and promotes a number of electrons to the next higher energy LL.  In the absence of impurities, no plateau can be formed, and the classical behaviour of the Hall resistance should be retrieved.
\item Upon adding, or subtracting more and more electrons away from the filled LL condition, one eventually saturates the number of impurity sites which localize electron states (or hole states)  and one finally  reaches a different plateau corresponding to a different filling factor.
\end{itemize}
In other words, the unprecedented accuracy of the IQHE plateaux is allowed by the very imperfections of the samples!

Another surprise was  mentioned above: in 2D, the transverse conductance of a Hall bar\footnote{Hall measurements are usually done on a long rectangular sample , whence the word ``bar''.} is independent of the geometric details of the sample! This is in sharp contrast with resistivity measurements of a 3D material, which are plagued by the accuracy constraints on length, diameter, etc.. Sample defects  as a cause of accuracy on the determination of a fundamental constant of physics: some would call this a paradox, some a contradiction. In any case this should stimulate some thoughts about the coexistence of contraries in the world, and in the process of its knowledge.

\subsection{Benefits of gauge symmetry.}\label{gauge}

The explanation of the IQHE does not require an explicit many-body wave function. The only necessary ingredients  are single electron wave functions in a magnetic field, and the exclusion principle for fermions, which states that no two electrons can occupy simultaneously the same state. In order to sort out the behaviour of one electron, one must chose a gauge for the potential vector, while formulating a  theory, the results of which must be independent of the gauge choice. To study the way electrons fill completely the lowest LL, the best gauge choice is that which makes theory simplest. In turn, this depends on the problem at hand. If one is interested in a multiply connected  disk geometry \cite{laughlin1}, the symmetric gauge comes in handy:  states are concentric rings. Laughlin found the first argument for the IQHE by resorting to such a geometry, and invoking gauge invariance. 

It is easier to describe the physics of the  Hall bar geometry by resorting to another gauge choice,  the Landau gauge, whereby the vector potential is perpendicular to the bar long dimension. This leads to  a rigorous  analogy between the conducting properties of the  electron liquid in a 2D Hall bar and a one dimensional  conductor. The conductivity of the latter is   a trivial undergraduate problem. But  the solution of this problem answers the main questions about the IQHE, and  allows to introduce a new concept in physics: that of the topological insulator.

\subsection{The topological insulator}\label{topological}

 A topological insulator is a conducting material the volume of which  has insulating properties\footnote{Inside the body, at least when disorder is sufficiently weak,  the excitation spectrum has a gap everywhere, which is the definition of an insulator at low temperature. The electron system is an incompressible liquid.}, while  conduction occurs on the  sample surface, with no dissipation in the QH filled LL case. The Hall bar has a zero resistance ``chiral'' one dimensional electron channel on each side of the bar. The current directions on either side of the bar are antiparallel.  The term ``chiral'',  means that electrons travel in one direction only \footnote{In a one dimensional conductor, electrons are allowed to travel in both directions.}, and cannot suffer back scattering (resistive) collisions\footnote{ A continuous one dimensional energy dispersion branch crosses the Fermi level close to the edges of the Hall bar.}. The vanishingly small longitudinal resistivity of a  macroscopic Hall bar is a macroscopic quantum phenomenon: a quantum phenomenon which occurs in the macroscopic world. 

In other words, the IQHE resistivity plateau, beside the paradox, or contradiction,  noted above about impurities, topples the traditional antinomy between insulator and conductor, in a different way  from that  of semi-conductors. Because it is both quantum and macroscopic, it is yet another example, besides superfluidity, ferromagnetism,  non locality effects in entangled macroscopic states, neutron stars, etc., of the fuzzy boundary between the quantum world and the classical one. 

 Classical behaviour is in general a property of macroscopic objects, at least in a wide range of experimental parameters. Macroscopic quantum phenomena compel physicists and philosophers to reflect upon, and clarify, the necessary conditions for emergence of the classical behaviour of matter from the quantum laws of the microscopic world. Topological insulators are a quickly developing subject nowadays which should have an impact on such studies.

\subsection{The Quantum Hall ferromagnet}\label{ferro}
Ferromagnetism is another   macroscopic quantum phenomenon the effects of  which have been known experimentally for a few thousand years. Iron (Fe), Cobalt (Co), Nickel (Ni), are transition metals which exhibit ferromagnetism at room temperature. Below a certain temperature, known as the critical temperature $T_c$, those metals, and their alloys, exhibit a continuous/discontinuous \footnote{The transition is both quantitatively continuous, and qualitatively discontinuous.}transition to a spontaneously broken symmetry state. The high temperature state has the full rotational symmetry of the Hamiltonian. At a temperature smaller than $T_c$, a macroscopic magnetization vector  appears, which selects an arbitrary direction in space. The symmetry of that state is reduced to axial symmetry around the magnetization vector. The historical development of the theoretical understanding of ferromagnetism has been slow in the twentieth century, with antagonistic schools claiming a thorough account of phenomena on the basis of conflicting hypothesis: one claimed that electrons in ferromagnetic metals are  localized at the lattice sites; this theoretical picture relied on the success of the so called Heisenberg Hamiltonian. The latter seemed at first to account for a number of ferromagnetism phenomena; it describes how localized electronic spins interact through inter-site coupling energies which change sign when the coupled spins change from parallel to antiparallel directions. The other theoretical picture paid more attention to the metallic character of ferromagnets, which is described by delocalized electrons, the wave functions of which spread in the whole sample volume, and interact through intra-site Coulomb repulsive energy. This school relied on the so called Hubbard Hamiltonian, which allowed to account for more detailed phenomena (alloy phenomenology, etc.).  Eventually, those antinomic pictures were tentatively reconciled in a complicated theory, whereby delocalized electrons interact in a way such that they behave much as localized ones. The delocalized picture has been able to explain many experimental features. The delocalized character is the dominant feature, the only which can reconcile magnetic, transport and alloy properties. But it is safe to say that the theory of magnetic transition metals is not yet in a fully satisfactory state. 

Not so for the Quantum Hall ferromagnet. It is the best known and understood  example of  ferromagnet. A two dimensional  Quantum Hall ferromagnet is stabilized because of Coulomb interactions between electrons  when the lowest Landau Level is full, and all others empty. The interaction energy is minimized when all spins are in the same  symmetric state, and the orbital part of the wave function is fully antisymmetric, as it should because of fermion statistics. This ensures that no two electrons can be close to one another. The many body wave function for the Quantum Hall ferromagnet is almost exactly known, from Lauglin's theory,  because there is a large gap for single reversed  spin  excitations, so that the whole Quantum Hall physics can be reduced to the study of a single  level. Furthermore, topological spin excitations of this ferromagnetic material, called skyrmions\footnote{Skyrmions are named after Skyrme who described similar excitations in nuclear matter.} exhibit both theoretically and experimentally the property that a spin texture carries necessarily with it a charge texture, so that a skyrmion carries one electronic charge. This is elegantly demonstrated theoretically by resorting to the Berry phase calculation of  quantum adiabatic  transport  around the skyrmion center \cite{sarma}. The main difference between the Quantum Hall ferromagnetism and that of, say, Fe, is -- besides the two dimensionality of the QH samples -- the negligible kinetic energy term in the Quantum Hall case: all electrons have the same kinetic energy, so the latter drops out of the picture: the competition (contradiction) between kinetic energy and interaction energy is entirely dominated by the interaction energy term. The latter is minimum when all spins are parallel.
%%%%%%%%%%%%%%%%%%%%%%%%%%%%%%%%%%%%%%%%%%%%%%%%%%%%%%%%%%%%%%%%%%%%%%%%%%%%%%%%%%%%%%%%%%%%%%%%%%%%%%%%%%%%%%%%%%%%%

\section{The Fractional Quantum Hall Effect}\label{FQH}
\subsection{The Laughlin theory}

The originality of Laughlin's work is manifold. He faced a problem for which, if the interaction energy were neglected, the ground state would have a huge degeneracy: that of a distribution of $\approx 3.10^{10}$ identical particles  among $10^{11}$  states of equal energy! The problem of finding the ground state with the interaction energy appears at first a lot more difficult than finding a needle in a haystack!

Laughlin   started with the fact  that the dominant energy, within the kinetic energy/potential energy conflict (contradiction) in the physical case at hand (two dimensional electron gas, partial occupation of the lowest LL) was the interaction energy energy; this prohibits using perturbation methods on a non-interacting ground state. He used a mathematical feature of  gauge symmetry (analyticity of wave functions in the LL)  to write a many-body wave function which is restricted to the lowest  LL. He used  the antisymmetry requirement for the fermionic wave function to find the general form of the many-body wave function such that particles would be equally distributed among all quantized surfaces $2\pi l_B^2$ in the sample surface. This was dictated by the failure of the Wigner Crystal proposal to account for the experiments \cite{FQHE}. He then proved that this unique state was separated in energy  from the first excited state energy by a large gap of order of $e^2/l_B$. This result has the force of a theorem. The numerical value of the gap is not exactly given by Laughlin's wave function, but an exact evaluation can only alter this in a minor quantitative way: the QH incompressible electronic liquid   is an absolutely true result of the theory, for the convenient  set of parameters  for which it has a lower energy than the Wigner Crystal. The integer or fractional value of the Hall conductivity plateaux in terms of the conductance quantum $e^2/h$ is also an exact result.

  The lowest Landau Level single particle eigenstates, in the symmetric gauge, are all proportional to an integer power $m$ of the two dimensional complex position  $z=x+iy$ ( apart from a gaussian normalizing multiplicative factor). This analyticity property is specific of the lowest LL. An analytic wave function contains no admixture of any component pertaining to  higher energy LL.  Two different single particle eigenstates in the LL differ by their respective exponents\footnote{The exponent $m$ is the integer eigenvalue of the angular momentum operator component along the direction of the magnetic field. }. There are as many values of $m$ as the degeneracy of the LL, which is, as we have seen, the number of flux quanta threading the surface sample. Laughlin's choice of mathematical formulation may be thought of as a beautiful combination of invariance (all eigenstates are analytic functions; they are all eigenstates of the angular momentum operator projection along the external field) and change (two single electron eigenstates differ by their center of coordinates $X,Y$ or by the quantized value of the angular momentum projection $m$  associated to $z^m= \left|z\right|^m\exp{im\phi}$).

 In order to grasp the beauty of the many body wave function for $N \approx 10^{11}$ particles  written down by Laughlin, it is useful to consider first the wave function for two particles. For a state which has rotation and translation invariance, the most general two particle antisymmetric wave function for particles with the same spin state is:
\begin{equation}\label{2p}
\psi^{(2)}(z,z') \propto \Sigma_m \alpha_m\left(\frac{(z-z')}{l_B}\right)^m
\end{equation}
modulo a gaussian factor which normalizes $\psi^{(2)}$ in an area of order $4\pi l_B^2$. $\alpha_m$ is the amplitude for the component with exponent $m$. 
Antisymmetry of fermion wave functions imposes that $m$ is odd, i.e. $m=2s+1$ where s is any integer. Strictly speaking, $\psi^{(2)}$  is multiplied by the symmetric spin ket $|++>$, which is left out, because spins play no role in the simplest theory described here.

It requires both deduction and  inductive intuition to admit that Laughlin's wave function  proposal for large $N$ is, logically:
\begin{equation}\label{s1} 
\psi^{(L)} (\left\{z_j \right\}) \propto \Pi_{i<j} \left( \frac{zi-zj}{l_B}  \right)^{2s +1}
\end{equation}
(where $s$ is a positive integer and  the gaussian normalizing factors have been left out for simplicity.)
In fact, for $s=0$ this function is precisely the full LL wave function with all electronic spins parallel: that of the IQH ferromagnet.

The wave function for the $1/3$ filling, and full spin polarization, of the lowest LL corresponds  to $s=1$.

This function has all the requisites of a many-fermions wave functions with all spins parallel: it is totally antisymmetric with respect to interchange of pairs of particles: the spin part of the wave function is symmetric since all spins are parallel.  This wave function vanishes when any two electrons have the same space coordinates.  This obvious remark has decisive consequences, because   the Coulomb repulsive energy is considerably reduced, since electrons are kept far apart from one another. Laughlin nailed the coffin of the Wigner Crystal  proposal (for this LL filling) by showing very simply that this state has lower energy than the Wigner Crystal. 

The astonishing property of this electronic  liquid (it is easy to see that it is a liquid, and  that translation symmetry is not broken by (\ref{s1})) is that its elementary excitations are vortices with fractional charge.  Charges of $1/3$ of an electronic charge have been predicted and experimentally confirmed using several techniques.

This discovery may inspire useful considerations on  reductionism; electrons and magnetic field are of course the basic bricks of the QH Effects. In the Integer QHE, one only needs solving the single particle Schr$\ddot{o}$dinger equation and finding  the quantum levels. However this does not help predicting the low temperature transport  properties of a fractionally filled Landau Level in a finite size sample such as a Hall bar. Reductionism is thus simultaneously correct, and wrong: The basic bricks are electrons and magnetic field, and the whole theory is derived from the Hamiltonian. Thus reductionism seems vindicated. But the quantum  state of a large number ($N\approx 10^{11}$ per $cm^2$) of electrons has qualitatively new features(ferromagnetism, skyrmions, coupling of charge and spin textures, excitations with fractional charge, etc.) which are out of reach of reductionist approaches, as emphasized in a general context by Nobel laureate P. W. Anderson, in his paper entitled {\it{ More is different}} (\cite{PWA2}). 

 The occurrence of fractionally charged particles in the Fractional  Quantum Hall states is an emergent property of the electronic system which cannot be predicted or understood, starting from the enigma of the fractional quantum Hall resistivity steps observed experimentally, except by recognizing, and working out,  the  new paradigm, i.e. the structure and properties of the strongly correlated many-electron system in two dimensions submitted to a magnetic field. A good proof of this is the failure of the Wigner Crystal proposal \cite{FPA}, based on electrons,  magnetic field, and the Hamiltonian, etc., to foresee what experiments revealed, a new state of matter. This paradigm illustrates another ubiquitous property of the world: the transformation of quantity in quality, in different ways: it required at first a sufficient lowering of the temperature, in a sufficiently intense magnetic fields to go over from the linear slope of the classical  Hall  resistivity with field to the quantum Hall plateaux.  Varying the magnetic field intensity allows to go over from an accurately quantized Hall plateau, with zero longitudinal resistivity, to a continuous variation of both (see figure \ref{fig02} ). And it took a sufficient density of 2D electrons in a sufficiently intense magnetic field to have a novel electron Quantum Hall  liquid with fractionally charged  excitations, which came as a complete surprise.

Laughlin's successful explanation of the IQHE and FQHE is based on a few fundamental and simple quantum mechanical  laws for a 2D electron system submitted to a static perpendicular magnetic field. Quantum mechanics needs nowadays no confirmation of its validity. If it did, the theoretical success it registers with the QHE would be an additional proof of the truth content of the Schr$\ddot{o}$dinger equation, of the superposition principle, of the motion of charged microscopic particles in a magnetic field, etc..

\subsection{Composite Fermions}\label{CF}

 The theoretical  discovery of  fractionally charged excitations in condensed matter physics, and their experimental observation has a general conceptual novelty which goes far beyond the apparently restricted area of the  physics at hand. It overcomes the belief that particles with fractional charge are only found in the strong interaction physics of quarks\footnote{In fact fractional charges had been proposed before in the theory of commensurate-incommensurate phase transitions.}.

Laughlin's wave function (equation \ref{s1}) explains the FQHE at LL filling fraction $\nu= 1/(2s+1)$. But it fails to account for a number of FQH plateaux, such as, for example $\nu= 2/5$,  which have been discovered after the first experimental findings. The plateau at  $2/5$  is part of the  series $\nu= p/(2sp +1)$, where $p$ and $s$ are positive integers. In 1989, Jain \cite{jain} proposed a trial wave function based on a re-interpretation of Laughlin's wave function. He simply separated the factors with exponent $2s+1$, as a product of two factors, one with exponent $2s$ and one with exponent $1$. The result is to interpret the FQH plateau at $\nu= 1/3$ as the fully occupied LL wave function for Composite Fermions, for the IQH plateau at an effective CF filling factor $\nu^*=1$ (instead of $\nu=1/3$ for $p=s=1$ for electrons). The other factor is interpreted as the wave function for $2s$ vortices attached to each electron. Each vortex is associated to a hole in the electron liquid, and carries  a flux quantum threading the sample. In other words, the Laughlin wave function for $\nu= 2s+1$ can be analyzed as that of a filled LL where each electron is attached to $2s$ flux quanta. An electron attached to an even number of flux quanta is a Composite Fermion (CF).

It is straightforward to check that  electrons attached to an even number of flux quanta, whatever the meaning or the mechanism of the attachment process, are fermions. The novelty introduced by Jain is to replace the wave function of the filled LL at $\nu^*=1$ (for $p=1$) by the wave function for $p$ filled Landau Levels\footnote{In addition this wave function has to be projected on the lowest Landau Level. This technicality is left aside in this paper.}. As a result, the plateau at $\nu=2/5$ is viewed as due to $\nu^*=2$ filled LL with CF carrying two flux quanta. The effective field applied to  the CF is the real field reduced by the density of flux quanta which have been attached to the electrons.

At this stage, CFs can be seen as a  trick to generalize Laughlin's theory of the FQHE at $\nu =1/(2S+1)$ to the new family of states with $\nu=p/(2sp+1)$ with $p>1$. Duhem \cite{duhem} would say CF help ``save appearances'', but has no nontological meaning. However the simple physical idea behind the CF story is that at fractional LL filling $1/(2s+1)$, the strong interaction energy  is minimized when electrons are kept on the average a distance $\sqrt{2s+1}l_B$ apart.

\subsection{The metallic state of Composite Fermions at $\nu=1/2$}. \label{metal}

Jain's CF picture was quite successful at interpreting various observed families of QH plateaux.
% But it could be (and was, probably still is ) thought of by many as a clever and convenient trick to ``account for phenomena'', with no ontological content.

The situation became somewhat more intriguing when it was remarked that if Jain's  construction is applied to the half filled Landau Level ( $\nu = 1/2$ ), the resulting effective field on Composite Fermions would vanish: indeed there is one flux quantum per state in the LL, so if two flux quanta are subtracted from the external field and  attached to each electron at half filling of the LL, this disposes of all available external flux quanta.  The prediction then was that the CF system at $\nu=1/2$ should be a metal, in zero apparent external magnetic field,  with a a filled Fermi surface (a Fermi disk  in 2D), the Fermi wave vector of which is predicted accurately. This was a daring proposal, if one remembers that we have an applied magnetic field on  a two dimensional electron system, the single electronic  levels of which are Landau Levels. 

Experiments, such as the measurement of CF cyclotron orbit, ultrasound absorption, etc.,  have beautifully confirmed this  prediction of the CF picture. The notion of flux quanta attached to electrons and subtracted from the real field to yield a vanishing  field at half LL filling  has gained flesh as a notion which may reflect, at least in some essential way, a real process of nature. Another daring proposal, taking the metallic state of CF at $\nu =1/2$ as a fact of nature is to propose that CF might condense in a BCS superconducting state \footnote{The pairing would not be the same as for ordinary superconductors with singlet spin Cooper pairs, but a different one, dubbed p-pairing. }, as happens for most ordinary metallic states in zero magnetic field at low enough temperature. This suggestion converges with a  theoretical pairing proposal which had been formulated earlier to account for the state at $\nu=5/2$. At the time of this writing, the author is not aware of experimental evidence for or against this spectacular suggestion which would, if confirmed, add a novel member  to the family of superconducting states.

CF, of course cannot exist in vacuum. If they are true entities, they need  the specific medium of 2D electron QH liquid, under magnetic field, to exist. The amount of correct predictions  and of  experimental results it allows to explain is impressive; future theoretical and experimental work should allow to clarify their ontological  status. 

%Whether Composite Fermions are bona fide particles or a theoretical framework to "`save the phenomena"' will be discussed in the section %\ref{sr} discussing "`Scientific Realism"'.

\section{Philosophy comments}\label{philo}

 In the above sections  I have  mentioned the existence of opposites which contradict each other within some objects in nature.

I have  stated a number of theoretical and experimental results as if they were  unquestionable. I have discussed  electrons as if I were sure they exist and their behaviour is truthfully described by quantum mechanics. Many a   philosopher of science dubbs electrons unobservable "`theoretical entities"' because one cannot see one with naked eyes. I have not questioned the existence of fractionally charged excitations in the FQH liquid, etc.. A number of philosophers of science persist in thinking  that the question of the reality of electrons, and a fortiori of fractionally charged excitations in nature is metaphysical.

What is the  rational attitude?
 
\subsection{Scientific realism}\label{sr}
In his interesting  book {\it{Representing and Intervening}}\cite{hacking}, Ian Hacking  reviews critically a number of positivist or agnostic philosophers, from Comte to Duhem \cite{duhem},  Carnap \cite{carnap}, Popper \cite{popper}, Kuhn \cite{kuhn}, Feyerabend \cite{feyer}, Lakatos \cite{lakatos}, van Fraassen \cite{frassen}, Goodman   \cite{goodman}, etc.. I define positivism here, loosely, as the   philosophical thesis  which reduces knowledge to establishing a correspondence  between theories, or mathematical symbols,  and phenomena, and deny that it may access ontological truths, dubbed "`metaphysics"'. Hacking  writes (p.131): "`{\it{ Incommensurability, transcendental nominalism, surrogates for truth, and styles of reasoning are the jargon of philosophers. They arise from contemplating the connection between theory and the world. All lead to an idealist cul-de-sac. None invites a healthy sense of reality...By attending only to knowledge as representation of nature, we wonder how we can  ever escape from representations and hook-up with the world. That way lies an idealism of which Berkeley is the spokesman. In our century (the twentieth) John Dewey has spoken sardonically of a spectator theory of knowledge...I agree with Dewey. I follow him in rejecting the false dichotomy between acting and thinking from which such idealism arises...Yet I do not think that the idea of knowledge as representation of the world is in itself the source of that evil. The harm comes from a single-minded obsession with representation and thinking and theory, at the expense of intervention and action and experiment}}"'.

I  agree with Hacking, inasmuch as he defends scientific realism. Scientific realism says that the entities, states and processes described by correct theories really do exist.  I do not underestimate, as I think Hacking seems to do, the explanatory power of a correct theory, such as Laughlin's theory for the Quantum Hall Effects. But  I believe that Hacking is fundamentally correct in stating that the criterion of reality is practice. He describes a technique which uses an electron beam  for a specific technical results. This convinces him that electrons exist. He thinks that "`{\it{reality has more to do with what we do in the world than with what we think about it.}}"`He discusses at length experiments, and points out that experimenting is much more than observing: it is acting on the world, it is a practical activity. The certainty I have about the reality of a QH incompressible electron liquid originates jointly from theory,  experiments, and  observations\footnote{Observation here is litteraly seeing a constant quantized value of a resistivity plateau on a screen or on a chart connected by electric leads to a macroscopic Hall bar.} of QH plateaux in $\rho_{xy}$ together with an exponentially low longitudinal resistivity $\rho_{xx}$. However this certainty is intimately connected, not only with the explanation provided by Laughlin, but also with various historical acquisitions of physics; for example the certainty that quantum mechanics describes correctly a vast amount of microscopic phenomena, which are at the basis of countless technologies which billions of humans use everyday,  which govern an increasing amount of industrial production, of all sorts of weaponry, as well as (devastating) financial operations worldwide\footnote{The classical Hall effect is used industrially in billions of computers.}. 

Hacking points out that not all experiments are loaded with theory, as Lakatos \cite{lakatos} would have it. The QHE belong to this category of discoveries  where experiments were intimately intertwined with theory. The actual initial specific theoretical prediction of the Wigner Crystal  turned out to be denied (falsified)  by the experimental result, but countless theoretical advances such as electromagnetism, the classical theory of the Hall effect,  Schr$\ddot{o}$dinger equation, band theory of cristalline material, etc., preceded the experiment and made it a rational endeavour. 

 Consider the following quotation: "`{\it{The question whether objective truth can be attributed to human thinking is not a question of theory but it is a practical question. Man must prove the truth -- i.e. the reality and power, the this-sidedness of his thinking --  in practice. The dispute over the reality or non reality of thinking that is isolated from practice is a purely scholastic question}}"'. That is Marx' thesis 2 on Feuerbach \cite{marxfeuer}! A second quotation is also relevant: "` {\it{The result of our action demonstrates the conformity ($\ddot{U}$bereinstimmung) of our perceptions  with the objective nature of the objects perceived}"'}. That is due to Engels\cite{engelspractice} who is also the author of a well known expression: "`{\it{The proof  of (the reality of)  the pudding is that you eat it}} "`. Compare with Hacking (p.146 of \cite{hacking}): "`{\it{"`Real"' is a concept we get from what we, as infants, could put in our mouth}}"'  

How come Hacking does not refer to those predecessors who have stressed, as he does, that the criterion of reality is practice?

The answer is probably in p.24 of \cite{hacking}: "`{\it{...realism has, historically, been mixed up with materialism, which, in one version, says everything that exists is built up out of tiny material blocks...The dialectical materialism of some orthodox Marxists gave many theoretical entities a very hard time. Lyssenko rejected Mendelian genetics because he doubted the reality of postulated genes}}"'.

It is a pity that Hacking, who reviews in many details various doctrines he eventually calls idealist, dismisses materialism on the basis of a version ("`{\it{of some orthodox Marxists}}"') long outdated. As for his dismissal of dialectic materialism, he is certainly right in condemning its dogmatic degeneracy during the Stalin era. But is this the end of the story?

Dialectical materialism  suffered a severe blow when it was used as official state philosophy in the USSR.  Much to  the contrary, nothing,  in the founding philosophical writings \cite{lenin,marx,engels} allowed to justify turning them into an official State philosophy.  This produced  such catastrophies as  the State support for Lyssenko's theories, based on the notion that genetics was a bourgeois science, while lamarckian concepts were defined at the government level as correct from the point of view of a caricature of dialectical materialism. It is understandable that such nonsense in the name of a philosophical thesis turned the latter into a questionable construction in the eyes of many.

% But a state judgment on  the validity of  a scientific result, on the basis of its conformity  with a simplistic interpretation of dialectical materialism,  was much %more the result of a political system devoid of elementary means of free public confrontation of ideas than due to 
Dialectic materialism itself is an open system, which has no lesson to teach beforehand about specific objects of knowledge, and insists \cite{engels} on taking into account all lessons taught by the advancement of science.

 It is perhaps time for a serious critical assessment of this philosophical thesis. The possibility of general theoretical statements  about the empirical world is not a negligible question. 

 Anyone who would dismiss Hacking's positions on realism,  on the basis that he made an erroneous statement about quantum mechanics \footnote{P. 25 of \cite{hacking}, one reads: "`{\it{Should we be realists about quantum mechanics? Should we realistically say that particles do have a definite although unknowable position and momentum?}}"'. In fact the very classical concept of trajectory, with simultaneously well defined  position and momentum is invalid for a microscopic particle. Particles do not have simultaneously definite position and momentum. Realism about quantum mechanics is justified, and natural as soon as one admits that classical behaviour is, in general, an approximation valid for actions large compared to $\hbar$.The non commutativity of the center of motion coordinates $X$ and $Y$ (equation \ref{XY}, discussed in section \ref{complem}) gives a realist example of the physical meaning of non commutativity of complementary operators. } would certainly not do  justice to his philosophical views. 

Hacking distinguishes between realism about theory and realism about "`entities"' (atoms, electrons, quarks, etc.) P. 26:"`{\it{The question about theories is whether they are true or are true-or-false...The question about entities is whether they exist }}"'. He is a realist about "`entities"' but doubts realism about theories. The QHE seem to be  a good example where realism about theory and about "`entities"' (electrons, incompressible liquids, skyrmions, anyons, fractional charges) are both relevant.

%In any case, is the distinction between realism about theories and realism about "`entities"' a fully rational one?

Dialectical materialism offers a different view. First, contrary to the caricature  mentioned above, materialism gives a clear answer to the "`{\it{gnoseological problem of the relationship between thought and existence, between sense-data and the world...Matter is that which, acting on our senses, produces sensations.}} This was written in 1908 by Lenin \cite{lenin}. It may look too simple when  technology (such as that used in QH experiments) is intercalated between matter (the 2D electrons in the Hall bar in figure \ref{fig01}) and the screens on which we read resistivity results. As discussed by Bachelard \cite{bachelard} and Hacking, a reliable laboratory apparatus is a phenomenon operator which transforms causal chains originating from the sample into readable signals on a chart or on a screen. Technology or not, matter is the external source of our sensations. So much for materialism. Dialectical materialism adds  a fundamental aspect of matter i.e. that contraries coexist and compete with each other within things in Nature. Depending on which dominates the competition (contradiction) under what conditions, the causal chains originating from the thing and causing phenomena will take different forms, which are reflected in theories. Epistemics and ontology are intimately intertwined. 

  Theories undergo a complex historical process of improving  representation  of how the things are. The Hall effects from Hall 1879 to this day are a good example. Some theories may prove false. Somme theories may be true.  But most  good theories are not, in general,  either true or false. Parameters and orders of magnitude have to be specified. 

\subsection{What about truth?}\label{truth}

Hacking  uses the following scheme, borrowed from Newton-Smith \cite{nsmith}, for scientific realism:
\begin{itemize}
\item 1. Scientific theories are either true  or false, and that which a given theory is, is in virtue of how the world is (ontological argument).
\item 2. If a theory is true, the theoretical terms of the theory denote theoretical entities which are causally responsible for the observable phenomena (causal argument).
\item 3. We can have warranted belief in theories or in entities, at least in principle (epistemological ingredient). 
\end{itemize}
 Hacking, in his book \cite{hacking}, "`{\it{drifts away from realism about theories and toward realism about entities we can use in experimental work}}"'. He stresses the role of practical activity to determine the reality of "`theoretical entities"'.

Item 2 above seems to correspond well to Laughlin's theory. So, according to item 1 it would seem Laughlin's theory is true and the Wigner Crystal theory \cite{FPA} is false. But it is not so. Laughlin's theory is also, in some respect, false: it does not explain the Jain Composite Fermion fractions. The Wigner Crystal theory, on the other hand  is true for Landau Level  occupations different from those for which the FQHE appears. And again it is false for the re-entrant QHE \cite{goerbig} where the electronic crystal seems to be an array of electron "`bubbles"' with integer number of electrons per bubble. In fact the compressible Wigner Crystal energy competes with the incompressible liquid energy. So Laughlin's theory is "`true"' in certain experimental conditions, inasmuch as it reflects and explicates the incompressibility of the electron liquid, and the QHE, for such conditions. The Wigner Crystal theory is "`false"' under those conditions, and "`true"' in others, when it accounts for the compressible insulating electronic crystal.

Nancy Cartwright \cite{cartwright} is a realist about entities, and an anti realist about theories. She denies that the laws of physics state the facts. She states that models used in physics and in applications are not true representation of how things are. Is this a serious criticism of models? As many modern philosophers of science, she rejects the correspondence approach to scientific truth: following the latter, a theory is true if it is a completely exact description of the thing. The correspondence approach is indeed  doomed to fail. The art of the physicist is to pick from reality the relevant parameters which are responsible for the causal powers of the thing  under study, at a given historical level of instrument technology, accuracy of measurements, etc.. This will define a model which abstracts away from reality, in general, a number of irrelevant elements. In that sense, models are not true representations of how things are. Better instruments, better accuracy of measurements  may allow to point out the effects  of factors which had been either neglected or  hitherto not  discovered. This allows to represent the thing in finer details. In general, the process of describing  how things are is an infinitely improvable one, which will  continue as long as humanity continues to explore the world. Stating that models are not a true representation of how things are is rather trivial, but it does not mean that truths about how things are  cannot  be obtained at a certain level of orders of magnitude, accuracy of measurements, etc.. How can Cartwright justify her skepticism when, in the case of QHE, the theory accounts for such observations as a resistivity plateau in terms of universal constants $e$ and $h$? Experiments and theory leave no room for improving on the exact value of the resistivity plateaux. Theory and experiment establish  rigorously  the existence of the incompressible QH liquid, etc.. So some truth about how things are has been obtained.  This question is also raised by Laughlin and Pines \cite{pineslaughlin} about the superconductivity  theory and the Josephson effect, or the Aharonov Bohm effect which give a number of theoretical results in terms of universal constants.

  A  representation need not be an exact reproduction of a thing to have a truth content. In general, in a given state of technology, it cannot be exact. The problem is with the dichotomy true/false. Kant \cite{kant} listed antinomies about which he said philosophy could say nothing, such as finitess/infinitess of the world, free will/universal causality, etc.. Hegel \cite{hegel} took up the question of antinomies and argued that contraries coexist, are ubiquitous  and compete with each other in human thought. Marx \cite{marx}   argued  that  Hegel's idealism is the reflection in human thinking  of  the coexistence and competition of contraries in the objective reality. 

Item 1 above in Newton-Smth's scheme is questionable because it does not take into account that, depending on practical conditions, on  technological progress, accuracy of apparatus, order of magnitude of parameters, etc., the mind independent reality (i.e. matter) may exert different causal powers. This is because in the competition (contradiction) between contraries, the dominating term, which  imposes the domination of certain causal powers at the expense of the dominated one may change and become the dominated one.

Was Fresnel's theory of light true or false? Was Newton's corpuscular theory of light false or true? Both were reconciled by quantum mechanics. Both were true in part, both were false in part. The microscopic world is simultaneously continuous (wave like in interference experiments) and discontinuous (particle like in the photoelectric effect). Depending on experimental conditions, one aspect dominates phenomena, or the other. Quantum mechanics shows that ``contraria sunt complementa'', as Bohr stated.

Hacking in 1985  breaks away from  idealist philosophies he criticizes for their lack of consideration of practice ("`intervening"') in the theory of knowledge. In my view, he remained in midstream. Realism is not materialism. Hacking quotes  d'Espagnat's realism (p.35) approvingly. This  betrays his ambiguity, for d'Espagnat \cite{espagnat} states that quantum mechanics proves that idealism has defeated materialism once for all, and  his realism admits the existence of God.  

  Are  there such  things as absolute truths? Yes seems to be the only possible answer: did the earth exist before mankind appeared and started to think about the world?   In that matter, I share, in part,  Boghossian's citicisms   \cite{boghossian} on the various brands of relativism and/or constructivism \cite{feyer,goodman,putnam,rorty,latour}. The epistemic system within which Quantum Hall plateaux exist and are analyzed in terms of universal constants  shapes the everyday life of a vast number of humans nowadays: no one can nowadays question the validity of  the Schr$\ddot{o}$dinger equation in the description of  the behaviour of electrons at the interface of semi-conductors. (Boghossian does not rely as much as he should on  the ``practice criterion'' of truth).
	%The Schr$\ddot{o}$dinger equation, or  is not a ``belief shared by a group of  scientists'', as Kuhn prtends

 The absolute truths I am referring to are truths which no rational person will ever be able to question.  They are also relative to a series of defined conditions, in particular a correct specification of the orders of magnitude of the various entities involved to establish it. In the case of the QHE, new conditions, such as  better accuracy of the  apparatus, different shapes or sizes of Hall bars, might reveal later on unsuspected new phenomena, but these will  not falsify the QH theory and the experiments in the defined set of reproducible conditions loosely described above.  A quantized value $\rho= e^2/h$ for the Hall conductivity plateau at $n=1$ is a rather good candidate for an absolute truth, given the experimental conditions described  in section \ref{prelude}. What I question here is the relevance of the dichotomy relative/absolute  with respect to truth, as well as the true/false dichotomy.

The history of physics proves that mankind, in a non linear, sometimes chaotic fashion, produces, in the sufficiently long run, a  process of accumulation\footnote{On the accumulation of knowledge, I agree with Popper \cite{popper} or Lakatos \cite{lakatos}.} of partial truths on the theories and entities of inanimate matter.  This historical process, in particular since the beginning of the industrial revolution,  triggers developments of  new tools of investigation, with better accuracy, with larger energies, larger pressures, larger magnetic fields, larger frequency ranges,  or lower temperatures, lower pressures, lower dimensions,  better controlled materials, better experimental resolution, and new practical applications, etc.. The hunger of private capitalist  investors for innovations which improve the labour productivity exerts a demanding  pressure in society for scientific results which have practical consequences (and a profitable market). New phenomena, in need of  new theories,   open new ways of approaching the processes of matter at such new scales of observation. Sometimes the new theories force to reconsider what was thought as established results \cite{kuhn}. But  the new theories, almost invariably, embody a number of ancient ones which describe previously known phenomena\footnote{One may even wonder if the main novelty in Copernicus theory, compared to the Ptolemy system of geocentric theory, was not to deny the Bible description of the Earth as the center of the world. Taking the sun as center of the planetary system certainly allowed simplification in the calculation of eclipses, but both  Ptolemy and Copernicus agreed that the problem was one of relative motion of the Sun and the Earth...}. For example, once the notion of a maximum velocity of signals is added to classical mechanics,  special relativity results, and  reduces to Newtonian dynamics when velocities are small compared to that of light; quantum mechanics reduces to classical mechanics when  the action is large compared to $\hbar$, etc..\footnote{ This statement does not have universal validity. It does not apply to macroscopic quantum phenomena, such as superconductivity,  the Quantum Hall Effects, etc.. }. 

 Kant's statements \cite{kant} on the epistemically  unreachable {\it{thing in itself }} seems to be true at any given time in history: knowledge, in general, at a given historical stage of humanity,   does not access all aspects of the ``thing'', which is what Cartwright perceives without placing it in  its historical evolving framework.  But contrary to Kant's  and Cartwright's thesis, there is,  in general,  no limit to the accuracy with which  epistemics approach  ontological truths.  At any finite time in human history, except in few cases, there is no complete  bridging  of the the gap  between the representation of the {\it{thing }} and the {\it{thing  in itself}}.  When topology is concerned \footnote{Topology deals with the study of shapes and topological spaces. Topological properties of space are preserved under continuous  deformations, such as bending, or stretching; this includes  connectedness, continuity and boundary. For instance, the statement : ``the Earth has the topology of a sphere in three dimensional space'' is an absolute truth. Many topological results are described by integer numbers, the accuracy of which is infinite: they represent  absolute  truths on a limited aspect of an  object.} absolute truths can be reached, but they are only partial truths about the ``thing''. However, for a materialist, there is also no  a priori limit to the accuracy with which the {\it{thing in itself}} can be known. In a given range of parameters, with a given accuracy of measurements -- i.e. given the suitable orders of magnitude --  there is no reason to question theories which are  overwhelmingly confirmed by practical applications involving the same orders of magnitude.   The reason why absolute truths of the sort I am discussing here may be reached is that there is no barrier between the inanimate material world, and humanity, which is itself a particular form of matter, and has evolved intricate individual and social means of investigating reality. Nobody  can prove that humans have no access to knowledge of some absolutely  true features of the world.
%%%%%%%%%%%%%%%%%%%%%%%%%%%%%%%%%%%%%%%%%%%%%%%%%%%%%%%%%%%%%%%%%%%%%%%%%%%%%%%%%%%%%%%%%%%%%%%%%%%%%%%%%%%

 Do we fully understand quantum mechanics? Perhaps not. While the debate is still raging in philosophical papers on  the superposition principle, on Bell inequalities, etc., would anybody be considered as behaving rationally if she tried to investigate the spectrum of the hydrogen atom, of the structure of the Mendeleieff table, of the low temperature specific heat of metals, of the structure  of molecules,  or the stability of atoms, of the laser, transistors, QH Effects, etc.,  without the help of the Schr$\ddot{o}$dinger equation? It may well be that the latter  be replaced one day by a better equation  which explains more things --for instance the reduction of the wave packet --, but this new equation will have to account for all the successes of present day text book quantum mechanics. This point of view is in some respects analogous to Chalmers' ``Non Figurative Realism''\cite{chalmers}. The latter claims, as I do, that there is no end to the progress of physics. However neither Chalmers, nor Hacking,   accept that reality can be self contradictory, and in particular that theories may access to partial truths, which are absolute truths when they are related to  their validity limits.

What I'm claiming here is that the theory of the QHE is now so well established, it has produced so many results, so many tests, so many applications, that if someone would spend time nowadays to find a theory based on different principles than those Laughlin described to account for the IQHE plateaux, or for the simplest FQHE ones, her intellectual capacities would be doubted by all experts,  on  rational grounds. I have regularly quoted the Nobel prize awards connected with the Quantum Hall experimental and theoretical discoveries. This is not due to some irrational cult for such awards. In my view, the process through which Nobel prizes in physics are awarded to individuals is a social one through which a  research advance is given a social reliability label. There are good reasons to believe that the Quantum Hall Effects exist, as they are reproduced and described in many published papers in scientific journals with  referees, and  a major understanding of their essential aspects  is provided by Laughlin's theory. The Nobel prize awards are but an additional reason to believe in both theory and experiments on QH Effects.

\subsection{Chalmers' Non Figurative Realism}\label{NFR}

Chalmers \cite{chalmers} has reviewed critically the works of Popper, Kuhn, Lakatos, and Feyerabend, and concludes that ``{\it{the aim of physics is to set limits for the application of present day theories and develop theories which are applicable to the world with a higher degree of accuracy}}''. His proposal is that of ``Non Figurative Realism'' (NFR): ``{\it{the world is what it is, independently of the knowledge we may have; our theories, inasmuch as they are applicable to the world, are always more than statements of correlations between series of observations}}''. However, his conclusion is that ``{\it{ NFR does not suppose that our theories describe entities within the world, such as wave functions and fields...We can evaluate our theories according to their degree of success in grasping some aspect of the world, but we cannot go beyond that, and evaluate their success in describing the world as it is really, because we have no access to the world independently of our theories...}}''.  Chalmers denies that physics may reach ontological truths.  He justifies his denial of the possibility of reaching truths by historical examples such as the fact that Newton's corpuscular theory of light cannnot be reconciled with the modern theory of light, or with the failure of Newton's dynamics when velocities  are comparable to that of light, etc.. He states that ``{\it{There is no concept of truth, the quest of which would be the ultimate goal of science }}''. 

In other words, Chalmers departs partially from the positivist attitude \`a la Duhem \cite{duhem}, following which theories establish a mere correspondence between mathematical signs and appearances, but cannot go beyond. On the basis of this, a fierce battle raged between energeticists and atomicists. The former denied all ontological truth to the concept of atom, while the latter stated that atoms exist in nature. History has given a clear judgment on this matter.  However, Chalmers agrees with Kant's transcendental idealism in denying that science can eventually access absolute truths about the world.

The weaknesses of Chalmers' position are: 1) his refusal of admitting the basic thesis of dialectical materialism, namely the statement of coexistence and struggle of opposites within the ``thing in itself'', 2) his under estimation of what Bachelard thought essential about physics: the notion of order of magnitude \cite{lecourt} and 3) his failure to recognize the criterion of practice as a criterion of reality. In this respect, Chalmers is one step behind Hacking. Both tend to have skeptic views about the relation of theory with truth. 
%The statements of physics about the world, to a great extent, rely on specifying the relevant  orders of magnitude  for the phenomena it accounts for, the accuracy of measurements, i.e. the order of magnitude of the parameters which are available to  technology, at a given moment in history.
  
 As for quantum mechanics, it may not be the ultimate theory of the microscopic world, but it is definitely true for all orders of magnitude available to day to humanity  for experiments or technology. Chalmers may be right when he  states that electrons are not the tiny individual charged massive spheres they were thought once to be, but the correct (non relativistic) theory of electrons definitely rests, as Putnam says \cite{putnam} on the definition of their rest mass and charge, the undiscernability principle,  and their description by Schr$\ddot{o}$dinger's equation\footnote{ Electrons in adequate conditions seem so far to be perfectly spherical \cite{sa}.}. In this paper,  I have given reasons to 	accept that the Laughlin  wave function  in two dimensions, under suitable magnetic field intensity is both an effective mathematical representation of the electron liquid, and reflects objective properties of that liquid, within the accuracy of present day measurements, the quality of present day Hall bars, etc.. In other words absolute truths have been stated, under relevant experimental conditions and orders of magnitudes: absolute truths are also relative ones. 
%%%%%%%%%%%%%%%%%%%%%%%%%%%%%%%%%%%%%%%%%%%%%%%%%%%%%%%%%%%%%%%%%%%%%%%%%%%%%%%%%%%%%%%%%

\subsection{Scientific pluralism}\label{pluralism}

What is at stake in a number of discussions above may be interpreted as the question of "`scientific pluralism"': Can it be that two or more conflicting theories account for the same phenomena, or account  for  the causal powers of the same objective real entities?

Dickson \cite{dickson} defines scientific pluralism as "`{\it{...the existence or toleration of a diversity  of theories, interpretations, or methodologies within science}}"'.

 Following Hillary Putnam \cite{putnam}, who discusses how things (e.g. electrons) are named, and how theories evolve, all sorts of incompatible accounts of the thing appear, all of which agree in describing various causal powers which may be employed while acting on nature. 

Cartwright \cite{cartwright} emphasizes that in several branches of quantum mechanics, searchers may use  a number of different models of the same phenomena. They can be mutually inconsistent, and none is the whole truth.

Dickson asks: {\it{"`How can one be a pluralist about science while respecting the (approximate) validity of our best scientific theories?"'}} In the course of his paper, he defends his own admission that as far as quantum mechanics is concerned, pluralism is justified and acknowledges the existence, and toleration of  a diversity of contradictory theories. He writes: {\it{I shall address  the most obvious and serious objection to such a view...namely, that it places the scientifically minded person in the intolerable position of explicitly endorsing contradictions within science (as a matter of principle and not merely as a pragmatic matter). To do so is to reject the scientific enterprise.}}

The authors in references \cite{hacking,cartwright,putnam,dickson} admit that there might be pluralism in theories, in various guises. But admitting  ontological contradictions within the thing seems to all authors, except perhaps Putnam, to be  prohibited  and intolerable.

Dickson describes various sorts of pluralisms (it seems one may define 27 types).  What I am defending here is that contradictions in epistemics may reflect  disagreements between erroneous vs correct theories, etc., but may also reflect\underline{ ontological contradictions}. One must free oneself from the secular aristotelian prohibition -- which does not mean dismissing aristotelian logic altogether, in its own domain of validity -- and face the possibility that accepting contradictions in nature is a way to save the scientific enterprise when science is pluralist. 

I do not agree with most things Dickson says about quantum mechanics\footnote{Such as stating that quantum mechanics has no dynamics. I cannot enter in details here about these disagreements.}.
%  \footnote{Quantum Mechanics has dynamics: Schr$\ddot{o}$dinger's equation depends on time and does not describe only probabilities, but also time dependent amplitudes. Classical dynamics, on the other hand, for more than two bodies, may have chaotic solutions, which are both deterministic and unpredictable, etc.. }; 
 I stress here that he, a philosopher convinced of the aristotelian prohibition of contradictions, is constrained to admit them  as a true feature of  theory.

  Laughlin's theory of the QHE, on the other hand,  seems to have  no theoretical rival, to be no  example of pluralism in theory. However, we have seen that the Wigner Crystal proposal \cite{FPA} is valid in certain magnetic field intensities  or electronic density ranges. This is due to a competition of two ground states within the free energy of the 2D electronic liquid. The analogy with the ice-liquid transition in water  helps understand that there is no complete exclusion of one dominated state from the dominant one: space and time dependent correlations reflecting the tendency to form the dominated phase generally exist in the dominant one. Contradictions in epistemics  may well reflect, at times, ontological contradictions. 

\subsection{What about the QHE?} \label{about}
The historical development of the QHE supports conflicting views of various science philosophers.
\begin{itemize}
\item Both Carnap \cite{carnap} and Popper \cite{popper} would find justification of their different approaches to knowledge in the development of the QHE physics. Carnap puts observations first: indeed, the QHE start with the observation of a Hall quantized plateau in $\rho_{xy}$ together with a zero of ${\rho_{xx}}$ (figure \ref{fig02}); but this observation was driven by theoretical expectations; for Carnap, verification to confirm  more general theories comes second. Popper would start from a theory, deduce consequences, then test to see if the theory is falsified. As shown in sections \ref{prelude}, \ref{FQH}, \ref{CF}, the various episodes of the QHE story give mixed examples of the two views. However, Popper's falsification thesis cannot account for the funding experiment \cite{IQHE}: even though it seemed to refute the Wigner Crystal proposal of reference \cite{FPA}, the latter proved to be correct for different filling fractions. 

 Even though Popper acknowledges the accumulation process of science, he claims that no theory ever accesses truth. A theory holds as long as it is not falsified. Popper would probably recognize that Laughlin's theory is a progress compared to ref.\cite{FPA}, but for him the incompressible QH liquid  only has  more verisimilitude than the Wigner Crystal proposal, until refutation. What has been described in sections \ref{prelude}, 
 \ref{integer} and \ref{FQH} does not allow this skeptical or agnostic attitude. The certainty we have about the existence of an incompressible QH liquid and the reproducibility of its empirical manifestations is a refutation of Popper's writings on what he dubbed the ``Hume induction problem''. 

\item Kuhn \cite{kuhn}, and Feyerabend \cite{feyer}, in contrast with Popper and Carnap, recognize the historical character of the evolution of theories. They do not clearly connect this history to the improvement of technology, of sample qualities, of measurement accuracies, contrary to the evidence described in section \ref{prelude}.  They deny that scientific knowledge may ever access truth. The QHE started with experimental surprises \cite{IQHE,FQHE}, which opened a phase of riddle solving theoretical activity, somewhat along the lines that Kuhn developed in his work on scientific revolutions \cite{kuhn}\footnote{The notion of scientific revolution had been developed some thirty years before Kuhn's writing by Bachelard, under the name of "`epistemological fracture"' ("` coupure \'epist\'emologique"' in French)}. However, contrary to Kuhn's ideas, the theoretical  revolution \cite{laughlin1,laughlin2} which followed  did not amount to replacement of ancient paradigms by new ones, and "`incommensurability"' with ancient paradigms: it resulted in the development of a new paradigm added to a vast body of unquestionable results. The QHE do not falsify the classical Hall effect, which is valid at metallic electron concentration and low magnetic field. This  confirms  Hacking's criticisms \cite{hacking}  about Kuhn's constructivist views.  The QHE deserves the name of revolution because it broke with theoretical methods previously used in many body problems, introduced new methods, novel concepts, novel theoretical entities (fractionally charged excitations, Composite Fermions, edge states, topological insulators, etc.) created a new field of physics, with new experimental and theoretical offsprings.
% However, this revolution does not conform with Kuhn's views, inasmuch as it did not refute a whole ancient field of physics, but created a new one %without falsifying previous theories, except one prediction,the Wigner Crystal. The latter is incorrect in a limited range of experimental %parameters, correct and confirmed in others.
\item Lakatos \cite{lakatos} does not agree with Carnap's and Popper's   theory/observation dichotomy, which is clearly irrelevant for the QHE. He also could claim that the QHE confirm his criticisms of Popper's fasification theory: the appearance of the first experimental QH plateau \cite{IQHE} seemed to  "`falsify"'the theoretical prediction by Fukuyama et al. \cite{FPA}, but did not lead to questionning the validity of quantum mechanics, of the interacting particles Hamiltonian, or the electronic theory of semiconductor interfaces, etc..
% The reason for this is in the huge  range of successful  theories, experiments and applications of the latter topics in scores of other fields. In %fact, the phenomena  of the QH plateaux, as stressed above, meant the emergence of laws of matter unsuspected until then. 
 This might appear as an illustration of Lakatos'views: he supported the notion of the growth of knowledge, and that of  "`protective belts"' around "`theoretical hard cores"' at the heart of research programmes.  In that respect the QHE might be considered as confirming Lakatos' views, and his criticisms of Kuhn's constructivism: quantum mechanics, electromagnetism, etc., were not abandonned because of the (temporary) failure of the Wigner Crystal hypothesis. 
\item The main common problem with Carnap, Popper, Kuhn, Feyerabend, Lakatos, Chalmers, Cartwright and Hacking is their  views about truth. They  hold that a theory is either true or false. This has been  discussed critically in section \ref{truth}. What the QHE story shows is that this dichotomy does not hold: the Wigner Crystal proposal is correct for certain LL filling fractions, incorrect for others where the QH incompressible liquid has lower energy. As is ubiquitous in Condensed Matter Physics, different ground states, with different symmetries, compete with each other: this  can be described as antagonistic orders within the interacting electronic system. One supersedes the other, depending on orders of magnitude of various parameters. The incompressible electronic liquid described by Laughlin's theory, within a specified range of experimental parameters, is a true theoretical entity, as proved by its account of experiments and confirmation of  predictions. 
%%%%%%%%%%%%%%%%%%%%%%%%%%%%%%%%%%%%%%%%%%%%%%%%%%%%%%%%%%%%%%%%%%%%%%%%%%%%%%%%%%%%%%%%%%%%%%%%%%%%%%%%%%%%%%%%%%%%
\end{itemize}

\subsection{Contradictions in the essence of things? Unity of opposites?}\label{contradict}

 The term ``essence'' used in the title of this subsection does not refer to the eternal nature of the ``thing'', as identified by its necessary properties, supposedly permanent through all its phenomenal manifestations. Essence here is understood as the effective set of causal chains
through which the ``thing'' has effects in the world, in space and time. Science has in general only a partial, but (in time) always improving access to this set.
 
 I have stressed a number of times that contradictions -- the coexistence and competition/struggle of opposites  --  appear fairly regularly, upon scrutiny, as intrinsic  features of reality. This notion, which is a fundamental aspect of dialectical materialism \cite{lenin,marx,engels,hegel,seve} is often tacitly considered as a superseded pathology of stalinist dogmatism.

The formulation by Engels \cite{engels} has been the basis of  the so-called  ``laws'' of dialectic materialism: the unity  and struggle of contraries (or equivalently, opposites),  the transformation of quantity into quality, and  the negation of the negation. The very term of ``law'' cannot be taken at face value, since it carries  with it an equivalence with scientific laws. Dialectical materialism is not a science, but its nature and its epistemic criteria  are closely related to science \cite{seve}.  The term ``thesis'' is probably more suitable.

A first observation is that dialectical materialism cannot be reduced, once and for all, to the body of the three ``laws'' mentioned above. Opposites, or contraries, which form a contradictory unity may be antagonistic (so that one of the poles struggles to destroy the other) or non antagonistic, so that the opposites coexist and interpenetrate each other \cite{seve}. This is an example of a development which occurred long after the formulation of the so-called laws. S\`eve  \cite{seve} points out the interest and necessity of  deepening  our understanding of the rich variety of forms of contradictions in nature, and of their development.

Another  observation is that the three  ``laws'' listed above have not been, to the best of my knowledge, of any conscious  use  in the development of the theory and experiments on the  QH Effects. This is not surprising, but has no consequence on the question of their relationships, and that of dialectical materialism, with the topic described in this paper.  Dialectical materialism does not provide a scientific methodology which would be valid at all times in all cases. It cannot assess beforehand the epistemic value of a given scientific approach of a specific phenomenon. 

Then after all, why bother?

 This is not a minor question. The history of physics is full of long battles between antagonistic views on scientific issues: geocentric versus heliocentric model; corpuscular or wave-like  nature of light; existence or non existence of atoms;  corpuscular or wave-like nature of electrons, atoms; existence or non existence of the ether; localized versus itinerant theory of magnetism, etc.. Sometimes, one theory takes the upper hand and  becomes the dominant one. Most times, new theoretical concept have appeared which blend the previously  contradicting views into reflecting an intrinsically contradicting reality: light is both corpuscular and wave-like, so are electrons, atoms, nuclei, etc.. Let me quote Bohr again: ``Contraria sunt complementa''.

Nowadays, thirty years after the discovery of the new ``high temperature'' superconductors\cite{bednorz}, the battle is still raging between followers of the ``strong interaction'' paradigm and the ``weak interaction'' one, between ``strong repulsive'' and ``weak attractive'' ones, "'electron-electron"' interactions and ``electron-phonon'' or "`electron-polarons"' ones, etc., with tens of thousands of published mutually conflicting papers in the scientific press: a good example of scientific pluralism, probably reflecting incompletely understood ontological contradictions\footnote{There is no space in this paper to discuss this, which will be the subject of forthcoming paper \cite{lederer-htcs}.}.

It is of interest to ascertain if a culture of contradictions, a culture about allowing the unity of opposites in our understanding of our scientific objects of studies is not one of definite interest and fecundity. 

The dominant ignorance, or rejection of the main thesis of dialectics in nature or in epistemics produces definite damages in  scientific discussions, as well as in  philosophical writings about physics.
% For example, renewing in 2009 the question ``particles or fields?'' about quantum field theory \cite{fraser} or denying  the ontological basis of the wave function for the benefits of a ``primitive ontology''\cite{allori} seems to reflect the permanence of an aristotelian  prohibition of ontological contradictions.
 This prohibition has proved its value in the development of logic; but the dogmatic exclusion of contradictions within nature does not seem to be  justified. Developing a culture which warns against unilateral views about the world seems an important task.

 One question in this paper is whether there is a way  to examine if the results I have described  give credence, or not, to a general philosophical statement  about the meaning of  dialectics  in nature. The issue is both about the dialectics of the thought process (the theory), and whether the latter reproduces, in its specific subjective way, correlations, relations and processes which exist objectively in nature, in the particular case of the physics I have discussed. The answer to this question, which this work suggests is positive, plays a  role in the discussion of the numerous contemporary  trends of subjectivist philosophies \cite{goodman,rorty,latour,putnam2}.

Even though, as I stated above, the three thesis of dialectical materialism as formulated by Engels are a somewhat schematic summary of the dialectics of nature, I proceed below to examine the QHE physics along those lines, for simplicity.
 
\begin{itemize}
 \item Starting with the negation of negation, I view it  as  a statement that all things evolve with  time, space,  or other parameters: change, or movement, for example, are based on the negation of, say,  position at time t, replaced by position at time $t+\delta t$, followed instantaneously by the negation of the new position, etc.. Is anything permanent in the world? Perhaps the total energy of the Universe, if this is a meaningful concept...But even if this be the case, the way  this energy is distributed in the various parts of the universe is a continuously evolving one.

In any case, beside its seemingly esoteric formulation, negation of negation seems a rather straightforward property of an ever changing world. Matter does not exist without some sort of motion. Motion is an example of continuous  negation of negation.  What about a  ground state of a many electron system such as described by the Laughlin wave function? Figure (\ref{fig02}) exhibits  the graph of the  Hall resistivity which may be \underline{thought of} as a series of negations of negations: just as any experimental curve of a quantity varying with some parameter: it is a succession of quantized plateaux of the Hall resistivity, with different values of the resistivity, separated by segments of   smooth variation of the same quantity. Here I have underlined that negation of negation appears also as a thought process, as a way of interpreting the experimental curve.

 What about a QH plateau? The resistivity value is not changed in a whole interval of the magnetic field; but when the field varies, the distribution of electronic states inside the sample  changes, since the degeneracy of the LL changes, so there is ``negation of negation'' of the electronic distribution even though the resistivity is pinned to a plateau quantum resistivity  value. Here change coexists with permanence, where the latter is expressed in terms of universal constants! Eventually this process leads to variation away  from (negation of) the  resistivity plateau.  What if the magnetic field is held fixed, and the state is in thermal equilibrium? A closer look at the condition for fixing the magnetic field shows that a constant intensity of the field is in fact an idealization of a field which has  intensity and spatial fluctuations sufficiently small that they can be neglected at the level of experimental accuracy. A ``fixed intensity of the field'' is a true property as long as the accuracy of the experiment allows to neglect its variations.  Fluctuations are as the oscillations of a pendulum: a continuous negation of  variation leading to a variation of opposite sign, and so on. Thermal equilibrium is also a continuous succession  of thermal exchange between two sources of heat such that a temperature increase is immediately countered by an opposite effect. 

Even though the concept of negation of negation has not helped anybody  at all in the discovery and explanation of the QH Effect, there seems to be no difficulty in interpreting things with it a posteriori. If one objects that, in the discussion above, dialectics is in the interpretation of phenomena, i.e. in the subjectivity of the author, she neglects the fact that this undeniable subjectivity seems convincingly to reflect  objective  processes.

\item The transformation of quantity in quality 

This has been mentioned at different times in this study, along with P. W. Anderson's paper \cite{PWA2} ({\it{More is different}}). This is probably the easiest thing to admit in the  dialectics of nature. Even though it seems trivial to many, it is at variance with the classical  Aristotle dichotomy  which opposes categories of quantity and quality.

In the QHE study, trivial examples abound, such as the transition from a classical behaviour of the Hall resistivity at low fields to that at high fields, where low, resp. high refer to the large, resp. small number of occupied LL. Another example is the qualitative change of the QH  Effects if disorder in the sample is low or high.
A less trivial example, and a spectacular one at that, is the transition of the 2D electron liquid under magnetic field from the Quantum Hall topological insulator at LL filling $\nu= p/(2ps+1))$ to a 2D metallic state at $\nu= 1/2$. A more trivial one  is that the QHE has no meaning for a few electrons: it only exists for a large enough number of electrons. 

We have seen along this work that this ubiquitous property of matter is important in discussing the validity of the reductionist analysis of phenomena. 

\item Contradictions, unity of opposites: the coexistence, struggle, and interpenetration  of opposites (or contraries) in Nature.

This is the most controversial aspects of dialectical materialism in the sciences of Nature. Many philosophers accept as straightforward  the notion that antagonistic classes exist within society, and that contradicting forces act at various levels  in the objects of social sciences. But what is the validity of the notion of contradictions within a thing in Nature? Can we give meaning to a statement that  contraries, or opposites, co exist, struggle  with  and interpenetrate each other within an object? On the other hand, how rational is the logical dogma following which  contradictions in nature are prohibited?

I have stated above that a standard attitude of physicists  is to determine what is the dominant parameter in a given object of study: entropy versus internal energy (thermodynamics), kinetic energy versus potential energy (classical and quantum mechanics, QHE), localization versus delocalization (wave functions, QHE), continuous versus discrete (quantum mechanics), symmetric versus asymmetric (phase transitions, QHE), reversibility versus irreversibility (thermodynamics, phase transitions, QHE) order versus disorder (phase transitions, QHE), long range versus short range (phase transitions, QHE), insulating versus conducting (thermal and electromagnetic properties of matter, QHE, topological insulators), interacting versus non interacting (ubiquitous, QHE),  magnetic versus superconducting (magnetism, superconductivity, QHE), attraction versus repulsion (molecules,  atoms,  nuclei, ) change versus invariance (phase transitions, mechanics, QHE, etc.). Etc.. 

Physicists will then examine (not in those words) under what conditions the dominant parameter will cease to be dominant and surrender its domination to the opposite pole: in general, this signals a qualitative change in the object under study.

The very list of those seemingly antinomic pairs, which clearly coexist, compete (struggle) and interpenetrate each other in various ways  is sufficient to convince that the "`third thesis"'  of dialectical materialism is meaningful  for a large number of problems discussed in physics, with a large diversity of ways for contradictions to appear.  Whoever thinks that such contradictions only exist in the subjectivity of human theoretical constructions will run into difficulties to prove her point, because of the many practical consequences which prove their objective validity. The contradictions I have listed above are convenient, at a given time in history,  to describe reality because they reflect, with more or less accuracy, contradictions which are at work in Nature.

If we think about the QH phenomena discussed in this paper, how can we avoid being impressed by the fact that a transverse resistance QH plateau, which coincides, within a range of  external field values, with a vanishing longitudinal resistivity, as well as a vanishing sample conductivity, is driven by the simultaneous coexistence of an extended conducting channel on the edge of the sample, with localized (insulating) states inside the sample? (A new contradiction appears here: inside versus outside...) I have stressed above the paradox (contradiction) between the accurate quantization of the Hall plateaux conductivity in terms of universal constants $e^2/h$,  and the contingency  of impurity disorder in the sample which causes  those plateaux to exist (here we have necessity versus contingency...). Notice also that depending on the amount of impurities, plateaux will vanish (too many impurities), appear, or disappear (no impurity disorder). We have here a rich variety of coexisting competing opposites. As the external magnetic field varies, the balance of localized states and extended edge  states shifts, until extended states within the sample connect the two QH bar edges;   extended states within the sample connecting one edge of the Hall bar with other  put an end to the (almost)   perfect conductivity of the   edge channel and the quantization of the transverse resistance plateau.

In all cases, depending of which pole, in the above list of opposites, dominates the other, and depending on the process which is at work in the contradiction,  the object within which such opposites coexist will have different properties. Depending on the way a given object is isolated, or submitted to external fields, the competition (or contradiction, struggle, etc.) between the opposites listed above may lead to qualitative differences depending of which pole dominates, the negation of negation will appear as an obvious process, or will not, etc..

\end{itemize}

What seems to emerge from the discussion above is that it is a posteriori justified in many instances to analyze a number of objects in nature in terms of dialectical materialist terms. The variation of their properties with various parameters leads to different ways for  opposites to combine within the object and  change its properties. This ``combination'' of opposites may be described as a ``struggle'', inasmuch as  the domination of one over the other determines specific properties. For example when kinetic energy dominates the potential energy in a mechanical system, the latter has a  qualitatively different behaviour than in the opposite case. Still in classical mechanics, a qualitative change of dynamics appears when the number of interacting bodies increases beyond $2$, and chaotic trajectories emerge while analytic trajectories continue to exist.  In quantum mechanics,  delocalization of the wave function lowers the kinetic energy;  the potential energy drives  localization. In thermodynamics, the internal energy drives order in a state of equilibrium;  the entropy drives disorder; order at short distance may coexist with disorder at long distance. In the QH Effects, when the orbital energy $\hbar \omega_c$ is small compared to the Fermi energy $\epsilon_F$ of the electron system in zero field, the behaviour is classical. When on the contrary $\epsilon_F << \hbar \omega_c$ the QH behaviour dominates. As explained earlier, the behaviour of the 2D electronic system under strong magnetic field is entirely dominated by the interaction energy, which is much larger than the width of a Landau Level. The latter is determined by the fluctuations of the impurity potential. The condition  on this width  for the QHE to appear is that it be much smaller than the inter LL energy $\hbar \omega_c$.

The fundamental question about dialectical materialism is whether this analysis in terms of struggle of  binary opposites is universally valid, and if such is the case, in what sense. In this paper I have discussed QH Effects. Their analysis  shows that it is plausible.  In various different ways,  dialectical materialism is supported by the QHE. This may sound trivial to many, and totally wrong to others. Attacking, with rational arguments,  the conventional dogma of the ``no contradiction principle'' in Nature following which  opposites cannot coexist in  a contradictory unity in Nature is not a negligible contribution to the philosophy of physics.

\section{Conclusion}\label{conclu}

The main points I have discussed in this paper about the QHE deal with:
\begin{itemize}
\item a) various contemporary  brands of philosophy of science.  I have spent  time with Scientific Realism (sections \ref{sr}, \ref{NFR}): Hacking, in particular, acknowledges practice as a central  criterion of assessing the reality of theoretical entities. I have discussed briefly, in subsection \ref{about}, the shortcomings of different authors'positions  more or less associated to positivism, who have not recognized the practice criterion.
All seem to agree on the difficulty, or impossibility of knowledge to access truths about the world, a skepticism about theories which   I have argued is impossible to hold on general grounds, and  for which, in sections \ref{progress}, \ref{revo}, \ref{history}, \ref{complem}, \ref{FQH},  the QHE   provide a counter example. 

\item b) scientific realism. The evidence for integer or fractional Hall conductivity plateaux in terms of the universal constant $e^2/h$, the existence of incompressible Hall liquids, as well as their competition with various forms of Wigner Crystals are established as true theoretical entities, as is the existence of fractional charges in the FQHE;  the QHE theories  contain undisputable elements of truth about the world. The status of Composite Fermions, discussed in subsection \ref{CF} is less clear:  the concept helps retrieving different families of QHE -- characterized by fractions $p/(2ps+1)$ where $p>1$ -- within the same category of theoretical entities described by the Laughlin wave function ($p=1$); the  prediction and observation of the metallic state at $\nu =1/2$ lends credence to the existence of entities  (CF) such that even numbers of flux quanta bind to electrons under determined physical conditions. The quantum dynamics of 2D electrons in a perpendicular magnetic field may well give rise to  this novel real entity. I am not aware of evidence which would prove their reality in terms of individual entities within the strongly correlated QH medium. Such evidence would be necessary to clarify their ontological status.

\item c) dialectical materialism. The ubiquitous transformation of quantity into quality is obvious in the QHE as it is in many other  processes of nature. It is worth mentioning that the material described in this paper is also an example of this in the process of knowledge: the accumulation of theoretical, experimental and technical advances of ``normal science'' discussed in section \ref{progress} eventually results in a revolutionary advance such as the QHE.  Can one draw conclusions about the unity of opposites within things in the case of the QHE?  I have argued in favour of a positive answer. This is connected with the transformation of quantity in quality: for example the quantitative change of magnetic field intensity governs the  transitions of QH resistivity plateau to  plateau, separated by qualitatively different intervals where the longitudinal resistivity is finite: this reflects the transition from  a QH regime of localized states within the Hall bar, and perfectly conducting chiral edge states  to a regime where both  edge currents, which have become resistive,  are connected by extended states within the Hall bar. This is an example where perfectly conducting edge states depend on localized states within the Hall bar, while resistive edge states are connected with conducting states between edges. Apart from the obvious fundamental  contradiction of quantum mechanical systems (waves versus particles), I have reviewed a number of contraries present in the QH system under magnetic field in section \ref{contradict}. The historical development of the QHE has been marked by the contradictory theories of the Wigner Crystal \cite{FPA} and the QHE incompressible liquid \cite{IQHE,FQHE}.  The dialectical combination of both is now a matter of consensus among physicists, and can also be viewed in the successful interpretation of the experiments revealing  re-entrance of the  QHE \cite{goerbig}. In fact both theories (QHE and Wigner Crystal or its variations) reflect true features of the electronic liquid in different ranges of electronic density or magnetic field intensity. Scientific pluralism is often, (not always) a reflection in theory of ontological contradictions in the thing. I have mentioned the long lasting, now superseded, struggle between the itinerant and the localized theory of ferromagnetism. The QHE offer yet another field of knowledge where the unity and  struggle of contraries within nature can be found in various forms.
\end{itemize}
%\subsection{Epistemics and ontology I}

In spite of their seemingly reduced area of  experimental physics, the Quantum Hall Effects  have led to a number of  advances of universal significance. This is perhaps due to the well defined set of parameters ($\hbar \omega_c, ~ e^2/l_B, ~ l^2_B$) which govern their appearance. There is little room in the physics of 2D electron liquids under magnetic field for other relevant parameters such as appear in other topics of condensed matter physics (ferromagnetism of $Fe, Co, Ni$ and alloys, High $T_c$ Superconductivity, for example). This is a striking effect of the contradiction between  the reduced specificity of the object of study and the universality of lessons one can learn from it. As far as lessons are concerned, this paper does not pretend to impress much the scientific practice in physics laboratories: it is spontaneously, though mostly unconsciously, materialist and dialectic. However I  hope it contributes to a larger interest of philosophical thinking for dialectic materialism.

\end{document}